\documentclass[a4paper,fleqn,usenatbib]{mnras}

\usepackage{mathptmx}
\usepackage[T1]{fontenc}
\usepackage{ae,aecompl}

\usepackage{graphicx}	% Including figure files
\usepackage{amsmath}	% Advanced maths commands
\usepackage{amssymb}	% Extra maths symbols

\title[CO emission from lensed DSFGs at $z>2$]{A survey of the cold molecular gas in gravitationally lensed star-forming galaxies at $z>2$}

\author[M. Aravena et al.]
{\parbox{\textwidth}{M.~Aravena$^1$\thanks{E-mail: manuel.aravenaa@mail.udp.cl}, J.~S.~Spilker$^2$, M.~Bethermin$^3$, M.~Bothwell$^4$, S.~C.~Chapman$^5$, 
C.~de Breuck$^3$, R.~M.~Furstenau$^6$, J.~G\'{o}nzalez-L\'{o}pez$^7$, T.~R.~Greve$^8$, K.~Litke$^2$, J.~Ma$^9$, 
M. Malkan$^{10}$, D.~P.~Marrone$^2$, E.~J.~Murphy$^{11}$, A.~Stark$^{12}$, M.~Strandet$^{13}$, J.~D.~Vieira$^6$, A.~Weiss$^{13}$, N. Welikala$^{14}$, G.~F.~Wong$^{15,16}$, J.~D.~Collier$^{15,16}$}
\vspace{0.5cm}\\ \parbox{\textwidth}{
$^1$ N\'ucleo de Astronom\'{\i}a, Facultad de Ingenier\'{\i}a, Universidad Diego Portales, Av. Ej\'ercito 441, Santiago, Chile\\
$^2$ Steward Observatory, University of Arizona, 933 North Cherry Avenue, Tucson, AZ 85721, USA\\
$^3$ European Southern Observatory, Karl Schwarzschild Stra\ss e 2, 85748 Garching, Germany \\
$^4$ Cavendish Laboratory, University of Cambridge, JJ Thompson Ave, Cambridge CB3 0HA, UK \\
$^5$ Dalhousie University, Halifax, Nova Scotia, Canada \\
$^6$ Department of Astronomy and Department of Physics, University of Illinois, 1002 West Green St., Urbana, IL 61801 \\
$^7$ Instituto de Astrof\'{i}sica, Facultad de F\'{i}sica, Pontificia Universidad Cat\'{o}lica de Chile, Av. Vicu\~{n}a Mackenna 4860, Santiago, Chile \\
$^8$ Department of Physics and Astronomy, University College London, Gower Street, London WC1E 6BT, UK \\
$^9$ Department of Astronomy, University of Florida, Gainesville, FL 32611, USA \\
$^{10}$ Department of Physics and Astronomy, University of California, Los Angeles, CA 90095-1547, USA \\
$^{11}$ Infrared Processing and Analysis Center, California Institute of Technology, MC 220-6, Pasadena, CA 91125, USA\\
$^{12}$ Harvard-Smithsonian Center for Astrophysics, 60 Garden Street, Cambridge, MA 02138, USA\\
$^{13}$ Max-Planck-Institut f\"{u}r Radioastronomie, Auf dem H\"{u}gel 69 D-53121 Bonn, Germany\\
$^{14}$ Department of Physics and Astrophysics, Oxford University, Denys Wilkinson Building, Keble Road, Oxford OX1 3RH, UK\\
$^{15}$ Western Sydney University, Locked Bag 1797, Penrith, NSW 2751, Australia\\
$^{16}$ CSIRO Astronomy \& Space Science, Australia Telescope National Facility, PO Box 76, Epping, NSW 2121, Australia
}}

\date{Accepted XXX. Received YYY; in original form ZZZ}

\pubyear{2015}

\begin{document}
\label{firstpage}
\pagerange{\pageref{firstpage}--\pageref{lastpage}}
\maketitle

\begin{abstract}
Using the Australia Telescope Compact Array (ATCA), we conducted a survey of CO $J=1-0$ and $J=2-1$ line emission towards strongly lensed high-redshift dusty star forming galaxies (DSFGs) previously discovered with the South Pole Telescope (SPT). Our sample comprises 17 sources that had CO-based spectroscopic redshifts obtained with the Atacama Large Millimeter/submillimeter Array (ALMA) and the Atacama Pathfinder Experiment (APEX). We detect all sources with known redshifts in either CO $J=1-0$ or $J=2-1$. Twelve sources are detected in the 7-mm continuum. The derived CO luminosities imply gas masses in the range $(0.5-11)\times10^{10}\ M_{\odot}$ and gas depletion timescales $t_{\rm dep}<200$ Myr, using a CO to gas mass conversion factor $\alpha_{\rm CO}=0.8\ M_\odot$ (K km s$^{-1}$ pc$^2$)$^{-1}$. Combining the CO luminosities and dust masses, along with a fixed gas-to-dust ratio, we derive $\alpha_{\rm CO}$ factors in the range $0.4-1.8\ M_\odot$ (K km s$^{-1}$ pc$^2$)$^{-1}$, similar to what is found in other starbursting systems. We find small scatter in $\alpha_{\rm CO}$ values within the sample, even though inherent variations in the spatial distribution of dust and gas in individual cases could bias the dust-based $\alpha_{\rm CO}$ estimates.  We find that lensing magnification factors based on the CO linewidth to luminosity relation ($\mu_{\rm CO}$) are highly unreliable, but particularly when $\mu<5$.  Finally, comparison of the gas and dynamical masses suggest that the average molecular gas fraction stays relatively constant at $z=2-5$ in the SPT DSFG sample.
\end{abstract}

\begin{keywords}
galaxies: evolution -- galaxies: formation -- cosmology: observations -- galaxies: starburst -- galaxies: high-redshift
\end{keywords}

%%%%%%%%%%%%%%%%%%%%%%%%%%%%%%%%%%%%%%%%%%%%%%%%%%

%%%%%%%%%%%%%%%%% BODY OF PAPER %%%%%%%%%%%%%%%%%%

\section{Introduction}

\begin{table*}
\centering
\caption{Observation Summary}
\begin{tabular}{lccccccc}
\hline
Source $^a$          & z $^b$           & Line  $^c$ & Freq. ($\nu_1,\nu_2$) $^d$              & Obs. dates  $^e$                               & Ph. Cal. $^f$ & Beam $^g$ & rms $^h$ \\
SPT-S$\ldots$               &                &                    & (GHz)                &                                                      &                          &            & (mJy) \\
\hline \hline
J011308-4617.7 & 4.2328  & CO(2--1)   & 41.800, 44.200 & 01, 02-Oct-2012                       & 0104-408	& $4.7''\times3.9''$, $81^\circ$ & 0.37 \\
J012506-4723.7 & 2.5148  & CO(1--0)   & 32.800, 37.500 & 03-Oct-2012			    & 0104-408	& $13.0''\times4.5''$, $89^\circ$ & 0.65 \\
J024307-4915.5 & 5.699    & CO(2--1)   & 34.500, 39.700 & 04, 07, 08-Oct-2012		    & 0252-549	& $6.2''\times5.1''$, $83^\circ$ & 0.22 \\
J034510-4725.6  & 4.2958  & CO(2--1)   & 41.500, 43.500 & 09-Oct-2012		 	    & 0332-403	& $6.2''\times4.7''$, $14^\circ$ & 0.80 \\
J034640-5204.9 & 5.6559  & CO(2--1)   & 34.500, 39.700 & 03, 04, 08, 09-Oct-2012          & 0332-403	& $6.7''\times4.9''$, $88^\circ$ & 0.35 \\
J041839-4751.8  & 4.2248  & CO(2--1)   & 41.800, 44.200 & 01, 02-Oct-2012			    & 0422-380	& $4.9''\times3.8''$, $85^\circ$ & 0.45 \\
J044143-4605.3 & 4.4771  & CO(2--1)   & 42.100, 43.800 & 03, 07-Apr-2013			    & 0454-463	& $5.1''\times3.9''$, $87^\circ$ & 0.34 \\
J045247-5018.6 & 2.0104  & CO(1--0)   & 36.600, 38.300 & 04, 06-Apr-2013			    & 0454-463	& $5.7''\times4.5''$, $87^\circ$ & 0.27 \\
J045912-5942.4 & 4.7993  & CO(2--1)   & 37.600, 39.700 & 05-Oct-2012			    & 0516-621	& $5.7''\times4.7''$, $78^\circ$ & 0.30 \\
J055155-4825.0 &  2.579   & CO(1--0)   & 31.000, 32.500 & 04-Apr-2013			    & 0537-441	& $7.0''\times5.4''$, $81^\circ$ & 0.40 \\
J210328-6032.6 & 4.4357  & CO(2--1)   & 41.400, 42.400 & 21, 23, 24-Mar-2013, 		    & 2204-540	& $5.4''\times5.1''$, $-81^\circ$ & 0.76 \\
J213242-5802.9 & 4.7677  & CO(2--1)   & 38.400, 40.200 & 13, 18, 19, 20-Sep-2013	    & 2052-474	& $5.5''\times4.4''$, $77^\circ$ & 0.33 \\
J213404-5013.2  & 2.7799  & CO(1--0)   & 31.000, 32.800 & 01, 02-Apr-2013			    & 2052-474	& $13.0''\times5.4''$, $53^\circ$ & 0.45 \\
J214654-5507.8 & 4.5672  & CO(2--1)   & 41.400, 42.400 & 22, 24, 29-Mar-2013		    & 2204-540	& $5.4''\times4.6''$, $89^\circ$ & 0.59 \\
J214720-5035.9 & 3.7602  & CO(2--1)   & 46.600, 48.400 & 28, 29, 30-Mar-2013		    & 2204-540	& $4.6''\times3.8''$, $88^\circ$ & 0.86 \\
\hline
\hline
\end{tabular}\\
\begin{flushleft}
\noindent $^a$ Source name.\\
\noindent $^b$ CO-based redshift obtained through millimetre spectroscopy with ALMA and APEX/Z-Spec in the case of SPT0551-48. \\
\noindent $^c$ Targeted low-J CO transition.\\
\noindent $^d$ Central observing frequencies for each of the two 2-GHz correlator windows, $\nu_1$ and $\nu_2$. One of them is tuned to observe the low-J CO, the other is used to have an extra measurement of continuum and to target a fainter molecular line.\\
\noindent $^e$ Observing dates.\\
\noindent $^f$ Phase calibrator used. In several cases, this coincided with the bandpass calibrator 0537-441. \\
\noindent $^g$ Synthesized beam size and position angle (PA).\\
\noindent $^h$ RMS level achieved at the low-J CO frequency in a channel of 50 km s$^{-1}$.\\
\end{flushleft}
\end{table*}

Panchromatic deep field observations have shown that the cosmic star formation rate density decreases by about an order of magnitude from $z\sim3$ to $z\sim0$ \citep[e.g.][]{lilly96, bouwens14, madau14}. An important contributor to the density of star formation at $z\sim1-3$ is an abundant population of galaxies with extreme star formation rates ranging from 100  to 1000 $M_\odot$ yr$^{-1}$ and large IR luminosities resulting from the re-processing of UV light by the large amounts of dust present. These galaxies have also been termed dusty star-forming galaxies (DSFGs) or submillimeter galaxies (SMGs), referring to the region of the spectrum in which some of the most luminous examples were initially discovered \citep[e.g.,][]{smail97,hughes98, bertoldi00, borys02}. Current observational evidence suggests they  are likely progenitors of local massive early type galaxies, and they appear to trace large galaxy overdensities at high-redshift \citep{brodwin08,viero09,daddi09a, aravena10, capak11,amblard11, toft14}. 

The primary reservoir of material in the interstellar medium (ISM) of galaxies is cold molecular gas. Large amounts of gas are necessary to sustain the large star formation rates (SFRs) in these objects to build up a massive ($>10^{10}$ M$_\odot$), luminous ($>L^\star$) galaxy in $<500$ Myr. The main components of the molecular ISM are H$_2$ and He with a minor fraction of heavier molecules. Given the difficulty to detect H$_2$ in the ISM, the CO molecule (the second most abundant molecule after H$_2$) has been commonly used to study the gaseous phase of the ISM in galaxies. In particular, the lowest rotational transition of CO $J=1-0$ represents the best studied tracer of the mass and spatial distribution of H$_2$ in galaxies \citep{omont07,carilli13}. The CO line emission can also be used to estimate the dynamical mass of the host galaxy, avoiding uncertainties due to differential dust obscuration and possible ionized outflows seen in the optical/near-infrared studies.

CO studies of DSFGs have often focused on the observation of high-J ($J\geq3$) transitions. However, such CO lines trace regions of enhanced star formation or active galactic nuclei (AGN) activity where high gas excitation is expected \citep{greve14}. 

Observations of low-J CO line emission in a few unlensed DSFGs at $z\sim1-3$ indicate that their gas masses can be $2\times$ larger and up to $3\times$ more spatially extended than expected based on $J\geq3$ CO transitions \citep[e.g.,][]{papadopoulos02,greve03,harris10,ivison11,riechers11b}. These results are reflected in the measured brightness temperature line ratios between the CO(3--2) and CO(1--0) transitions, $R_{31}=T_{32}/T_{10}$. For local thermodynamic equilibrium (LTE), this ratio is expected to be close to unity, however variations of the line ratios in individual sources make predictions of the low-J CO transitions unreliable. Most CO(1--0) line measurements in distant galaxies, including DSFGs and relatively quiescent star-forming disk galaxies at high-redshift, indicate that $R_{31}$ may be frequently as low as 0.5 with significant source-to-source variations, exemplifying the fact that $J>2$ CO emission lines may not necessarily trace the whole extent of the molecular gas or their total dynamical masses in most cases \citep[e.g., ][]{dannerbauer09, aravena10, carilli10, swinbank10, harris10, ivison11, danielson11, thomson12, bothwell13, sharon13, sharon15, aravena14, hodge15, bolatto15}. Indeed, CO measurements in different samples of DSFGs show a significant dispersion, with a median $R_{31}=0.52\pm0.09$ found by \citet{bothwell13} and a mean $R_{31}=0.86\pm0.08$ by \citet{spilker14}. In the case of luminous high-redshift quasars, the CO emission is found to be in LTE typically out to $J=5$ and the distribution of CO emission, including low-J CO, is mostly found to be concentrated in the inner kpc of the host galaxy, although there is evidence that in some cases the CO emission can be aligned with the radio jet axis \citep[e.g.,][]{papadopoulos08, elbaz09, emonts14}. Thus, measurements of molecular gas masses based on $J\geq3$ CO transitions become difficult due to the uncertainty in the unknown line ratios.

Since observing low-J CO emission at high-redshift is observationally time-consuming, current CO studies have focused on the most luminous galaxies or gravitationally lensed objects. Methods have been proposed to measure the gas masses from dust mass determinations based on far-IR and submillimeter continuum observations \citep{bloemen90, israel97, dame01, leroy11, magdis11, magnelli12, scoville14}. This could open an important window to study the fainter galaxies as continuum observations typically require less observing time, particularly in the era of the Atacama Large Millimeter/submillimeter Array (ALMA). Ideally, both molecular line and dust observations are required because they can provide independent estimates and  together yield tighter constraints on the ISM conditions.

The discovery of a population of rare (n$\sim$0.1 deg$^{-2}$) and extremely bright ($S_{\rm 1.4mm} > 15$ mJy) mm-selected galaxies in a deep, multi-band survey over $\sim2500$ deg$^2$ of sky with the South Pole Telescope \citep[SPT; ][]{carlstrom11}, or in the several hundred square degree far-infrared surveys carried out with the {\it Herschel Space Observatory} \citep{pilbratt10}, allows for detailed studies of the ISM properties of DSFGs out to the highest redshifts \citep{negrello10,vieira10,wardlow13,mocanu13}. Follow-up high-resolution 870$\mu$m observations with ALMA of a sample of SPT sources showed these are strongly gravitationally lensed galaxies at high-redshift \citep{hezaveh13,vieira13, weiss13}. Similar observations with the Submillimeter Array in a sample of bright Herschel and Planck sources recently found similar results \citep{bussmann13, canameras15}.

After the discovery of bright millimetre SPT sources in a systematic and unbiased way, a major quest began in order to characterise their properties and understand their nature. 
Blind CO spectroscopic observations with ALMA enabled the redshift confirmation of 28 SPT DSFGs, including two of the highest redshift DSFGs in the literature \citep{weiss13}. This represents an ideal sample to study the conditions of the ISM in distant galaxies, with most of the galaxies in the SPT sample ($\sim$70\%) confirmed at $z > 2$. Thus all the ALMA detected CO transitions in the 3-mm band are $J\geq3$ \citep{weiss13}. Observations of the low-J CO emission are thus necessary to directly trace their molecular gas content and further investigate the nature of these galaxies .

The SPT DSFGs were shown to have large IR luminosities, being $>10^{12}\ L_\odot$ for most sources even after correcting for lensing magnification \citep[Spilker et al. in prep; ][; This work]{hezaveh13,ma15}. Stellar mass measurements of a few SPT DSFGs indicates that they are located above the predicted sequence of secularly evolving star-forming galaxies at $z\sim2-3$ in the stellar mass versus SFR plane \citep[e.g.,][]{daddi07, elbaz07, noeske07, pannella09, peng10, karim11, rodighiero11}. The high specific star formation rates implied therefore suggest that SPT DSFGs are living in an active starburst phase, likely triggered by major mergers.

In this paper, we present a systematic survey of the low-J CO line emission ($J=$1--0 and 2--1) in a sample of gravitationally lensed SPT DSFGs at $z=2-6$, conducted with the Australia Telescope Compact Array (ATCA). 
In Section \ref{sect:data}, we present our source sample, the ATCA CO dataset, lensing models and dust properties that will be used throughout in this study. In Section \ref{sect:results}, we show the results obtained from the ATCA observations. In Section \ref{sect:analysis}, we derive physical properties of the SPT DSFGs based on the reported CO line observations (magnifications, sizes, masses). In Section \ref{sect:discussion}, we discuss the possible implications of our results. Finally, Section \ref{sect:conclusions} summarizes the main conclusions from this paper. Hereafter, we adopt a standard $\Lambda$CDM cosmology with $H_0=71$ km s$^{-1}$ Mpc$^{-1}$, $\Omega_\mathrm{M}=0.27$, and $\Omega_\Lambda=0.73$.

\begin{figure*}
\centering
\vspace{3mm}
\includegraphics[scale=0.33]{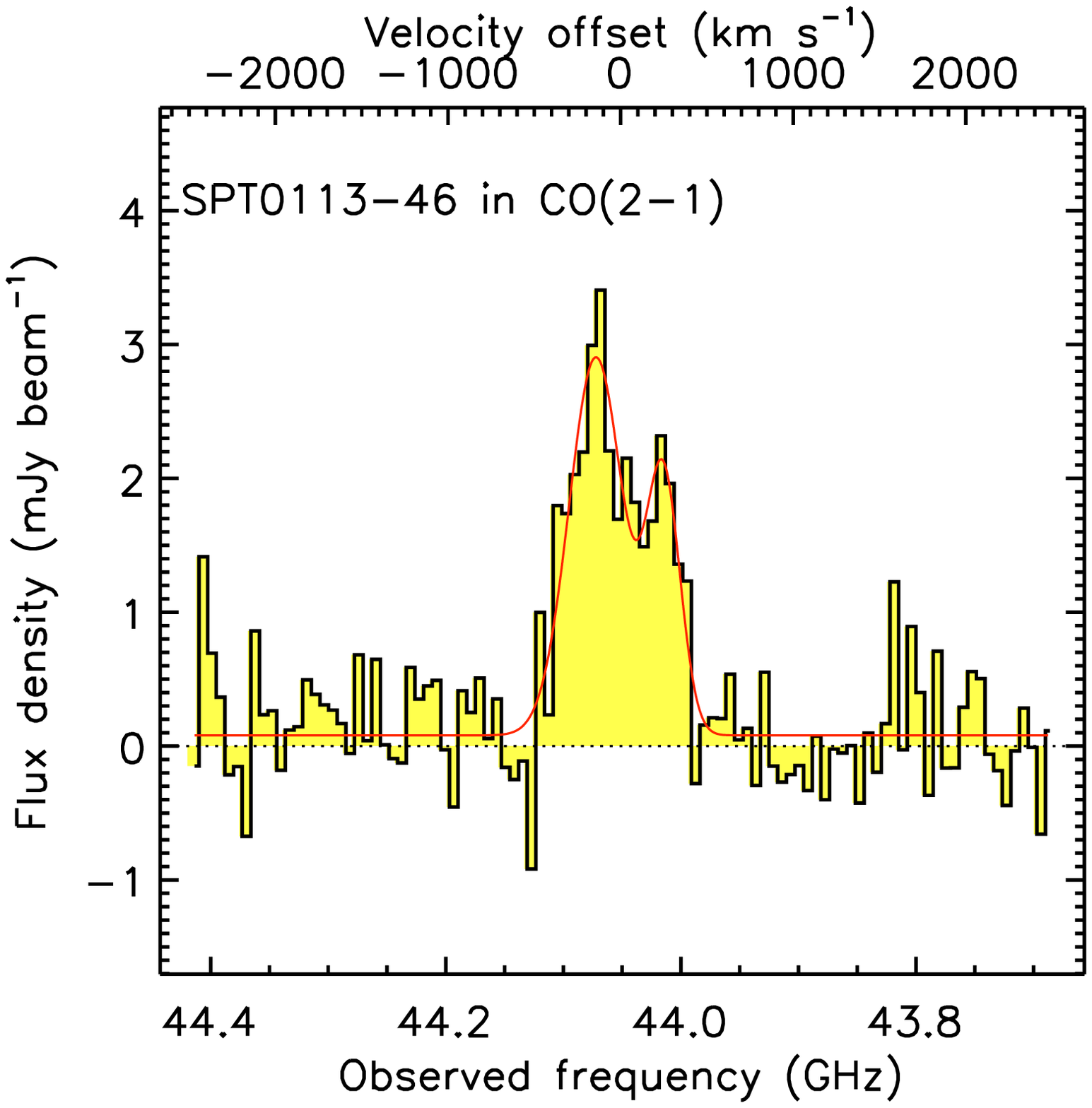}
\includegraphics[scale=0.33]{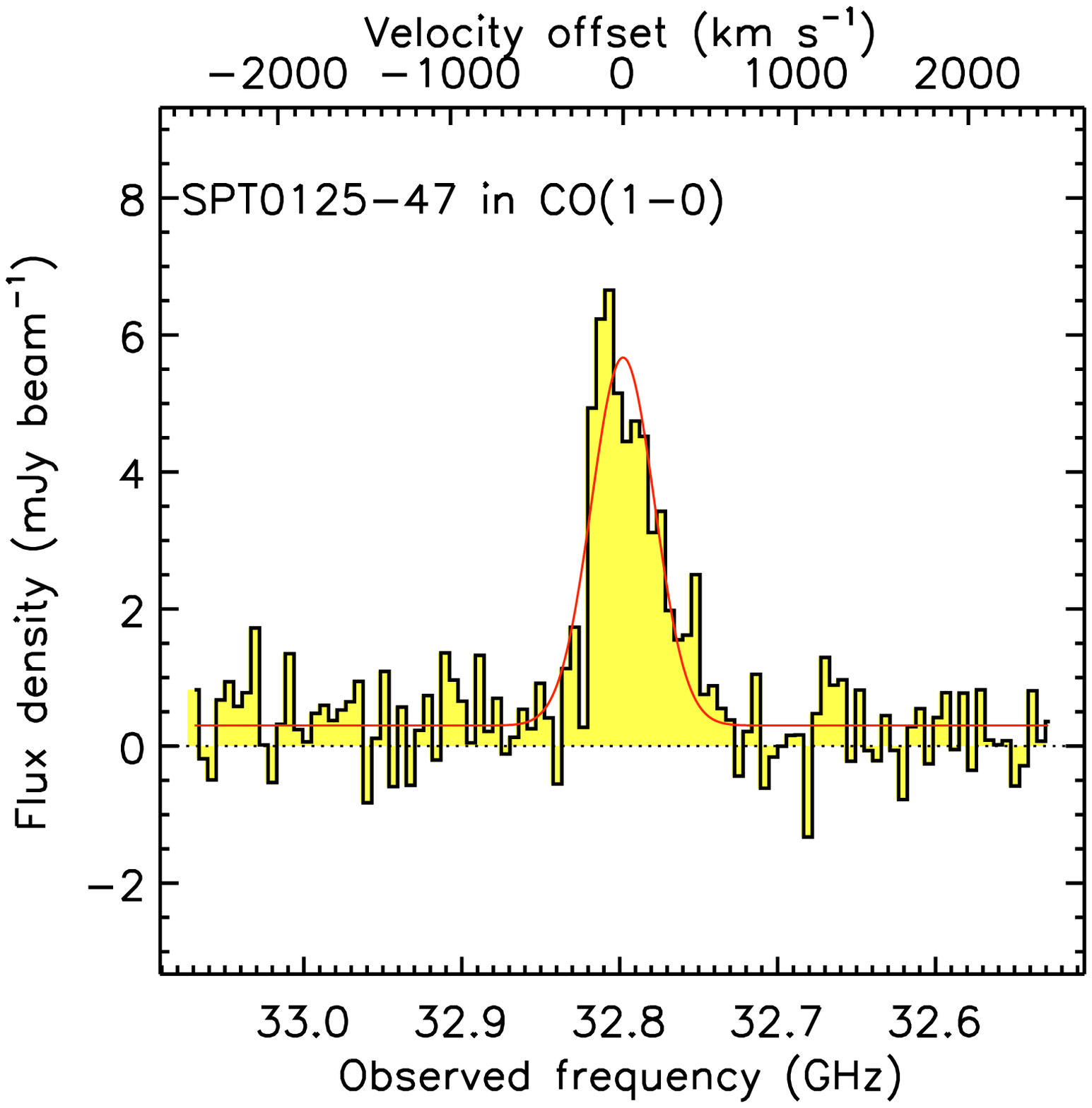}
\includegraphics[scale=0.33]{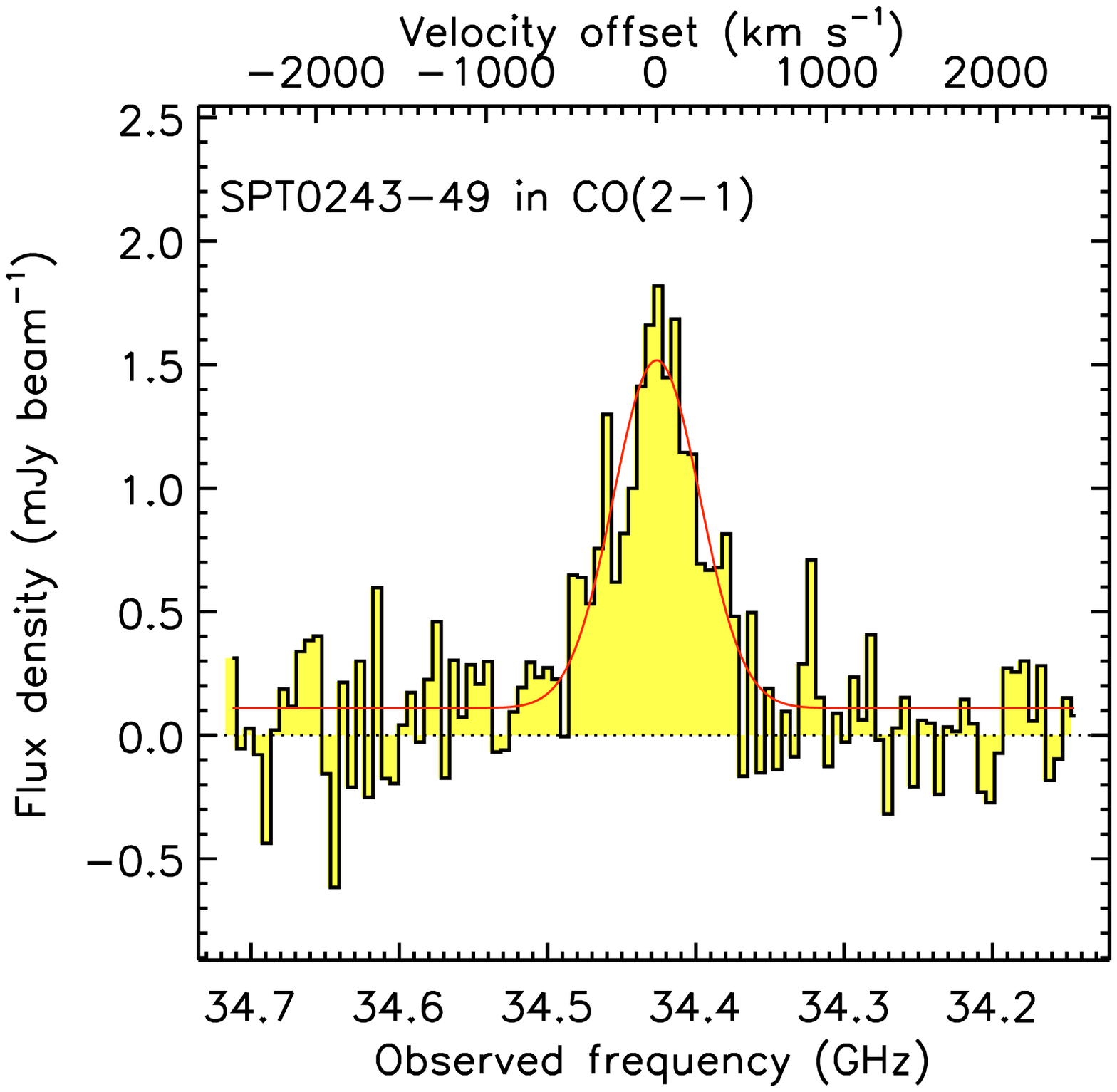}\\
\vspace{6mm}
\includegraphics[scale=0.33]{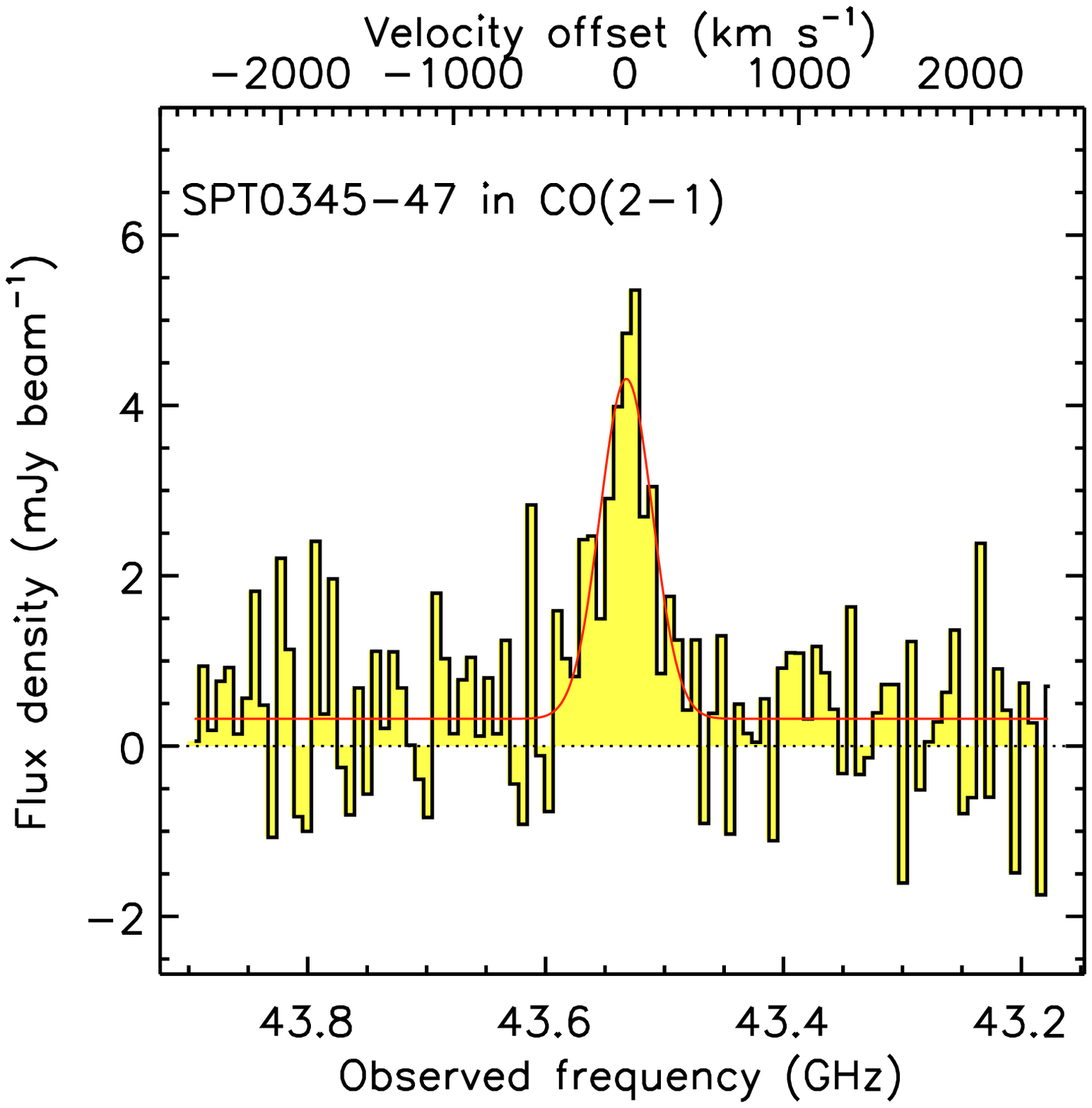}
\includegraphics[scale=0.33]{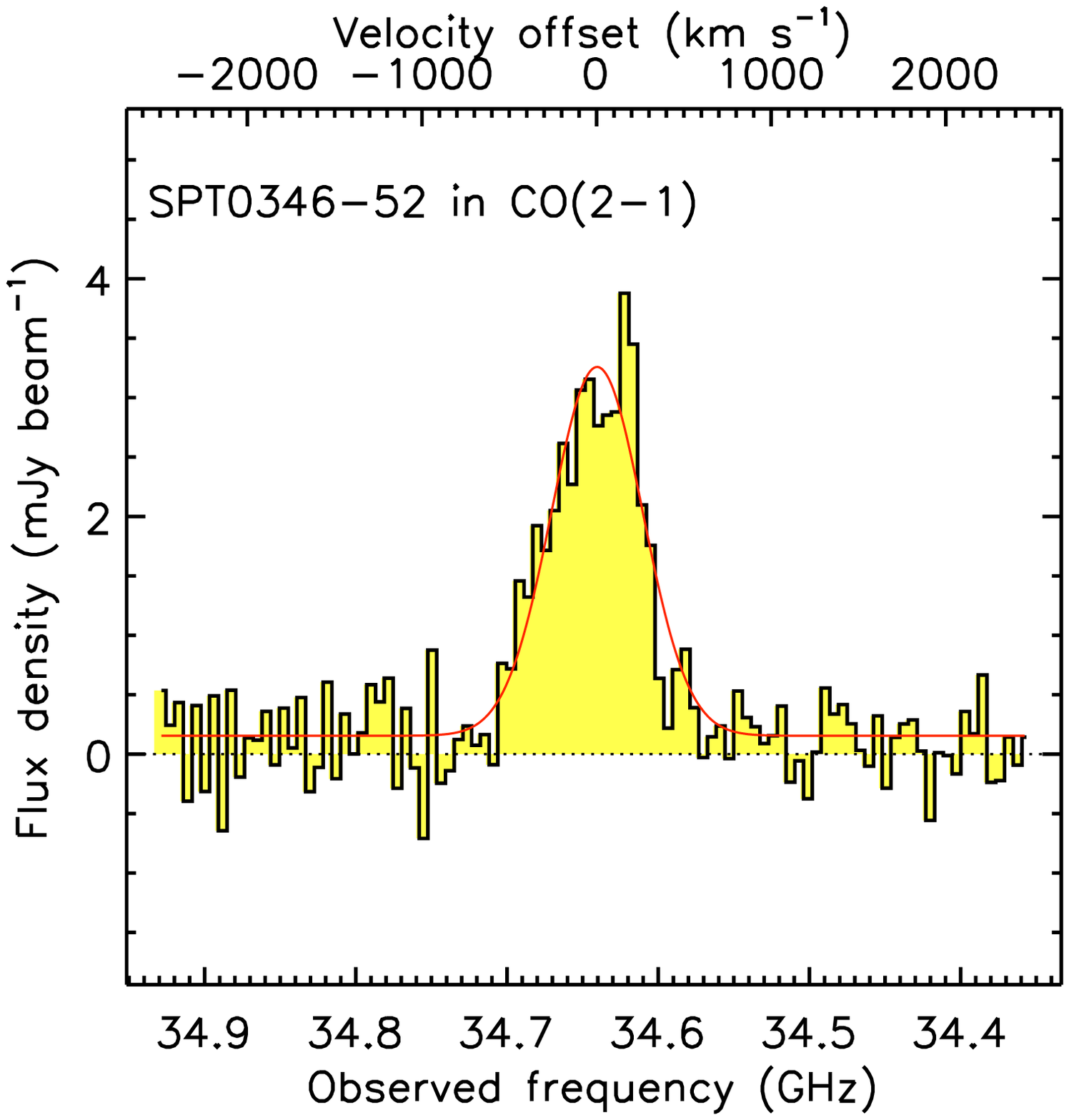}
\includegraphics[scale=0.33]{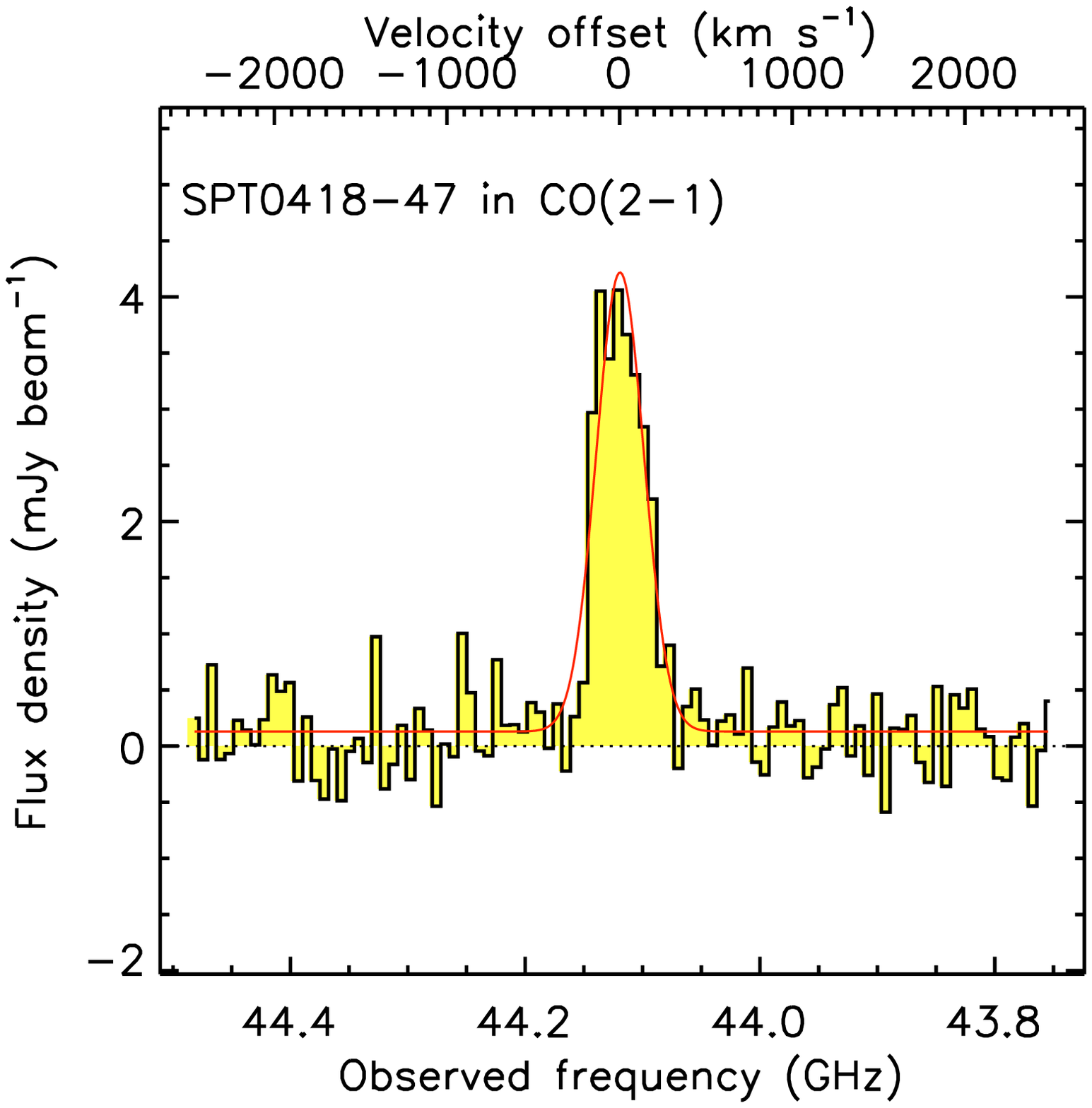}\\
\vspace{6mm}
\includegraphics[scale=0.33]{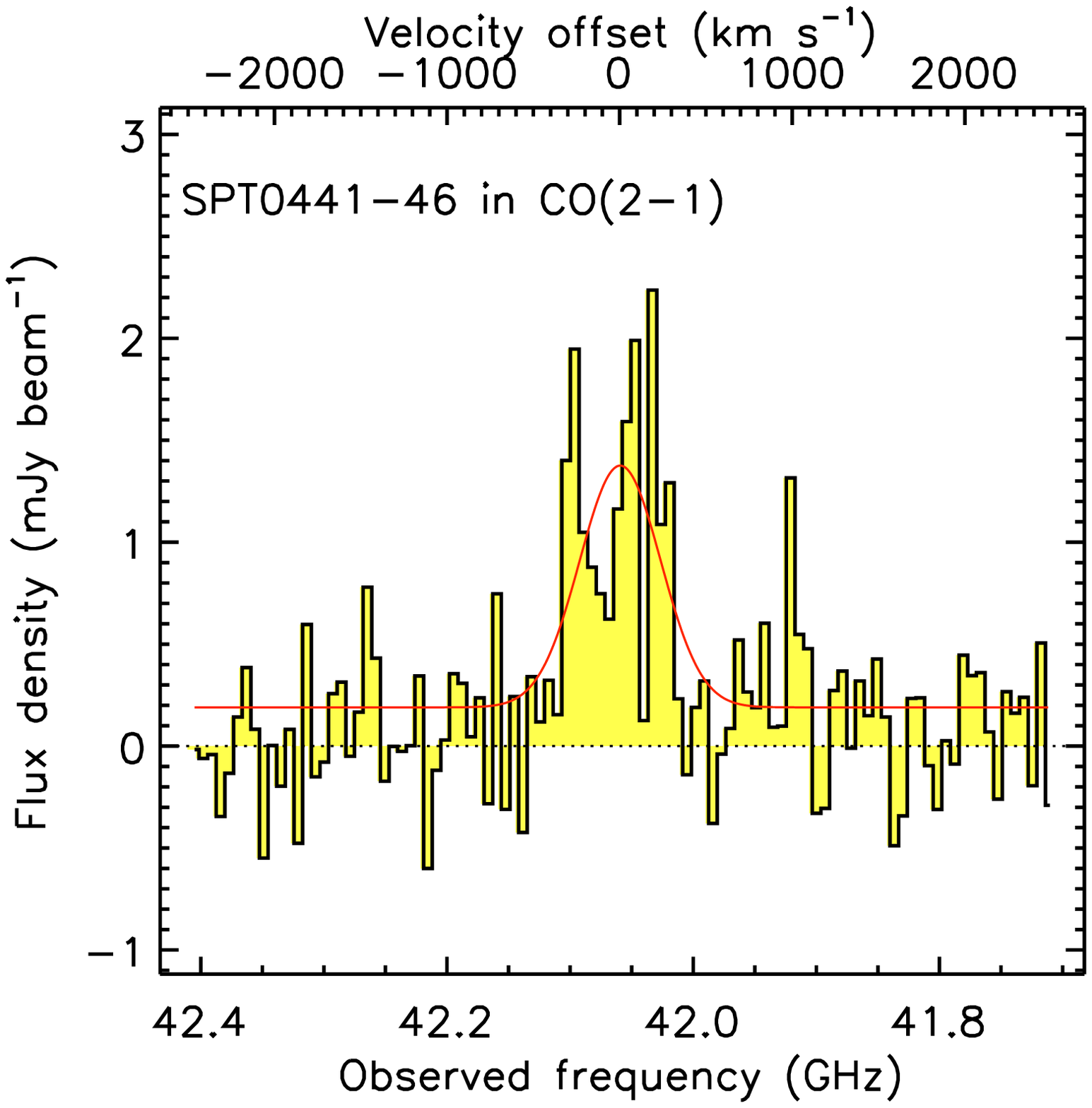}
\includegraphics[scale=0.33]{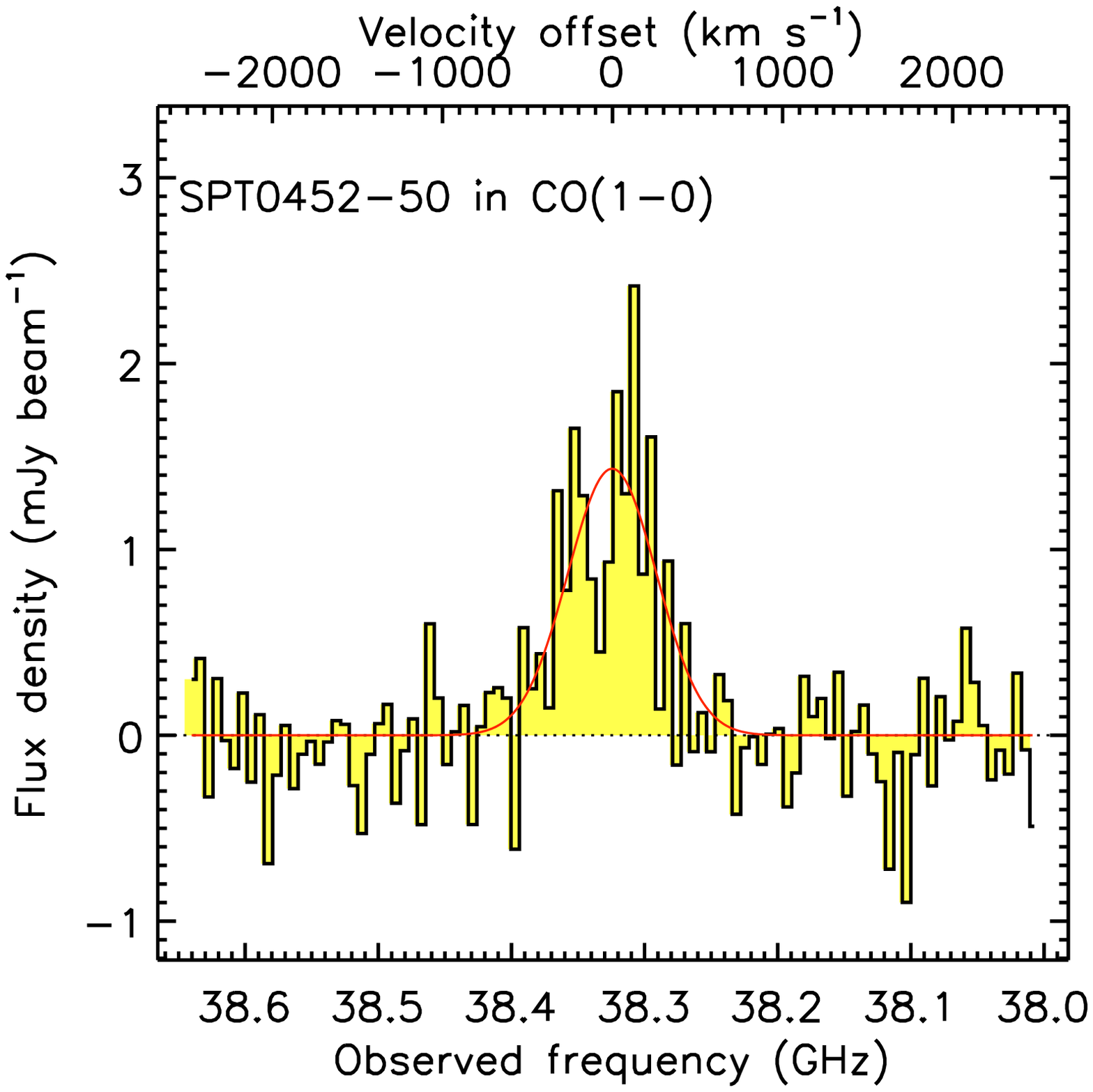}
\includegraphics[scale=0.33]{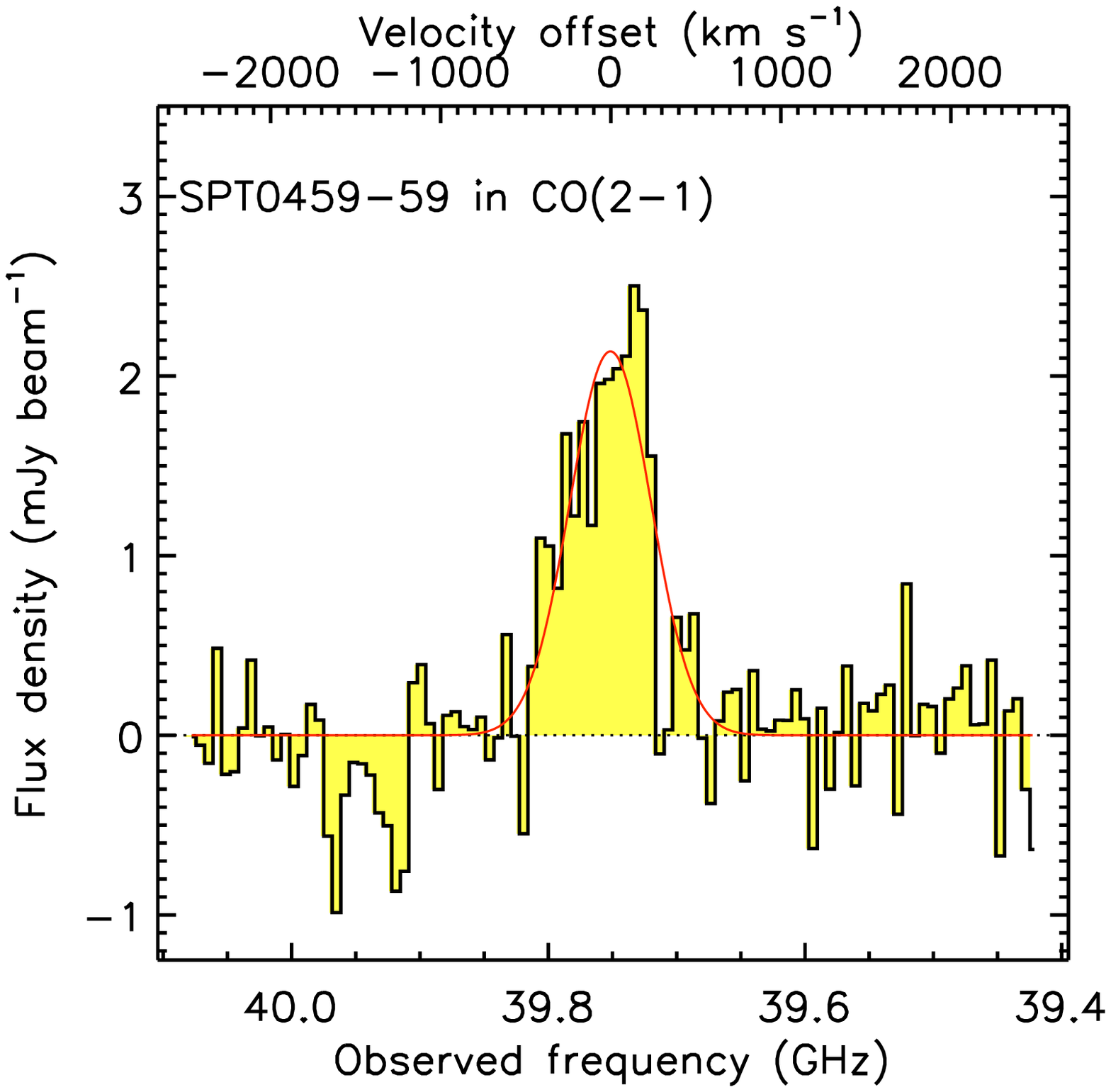}\\
\vspace{6mm}
\includegraphics[scale=0.33]{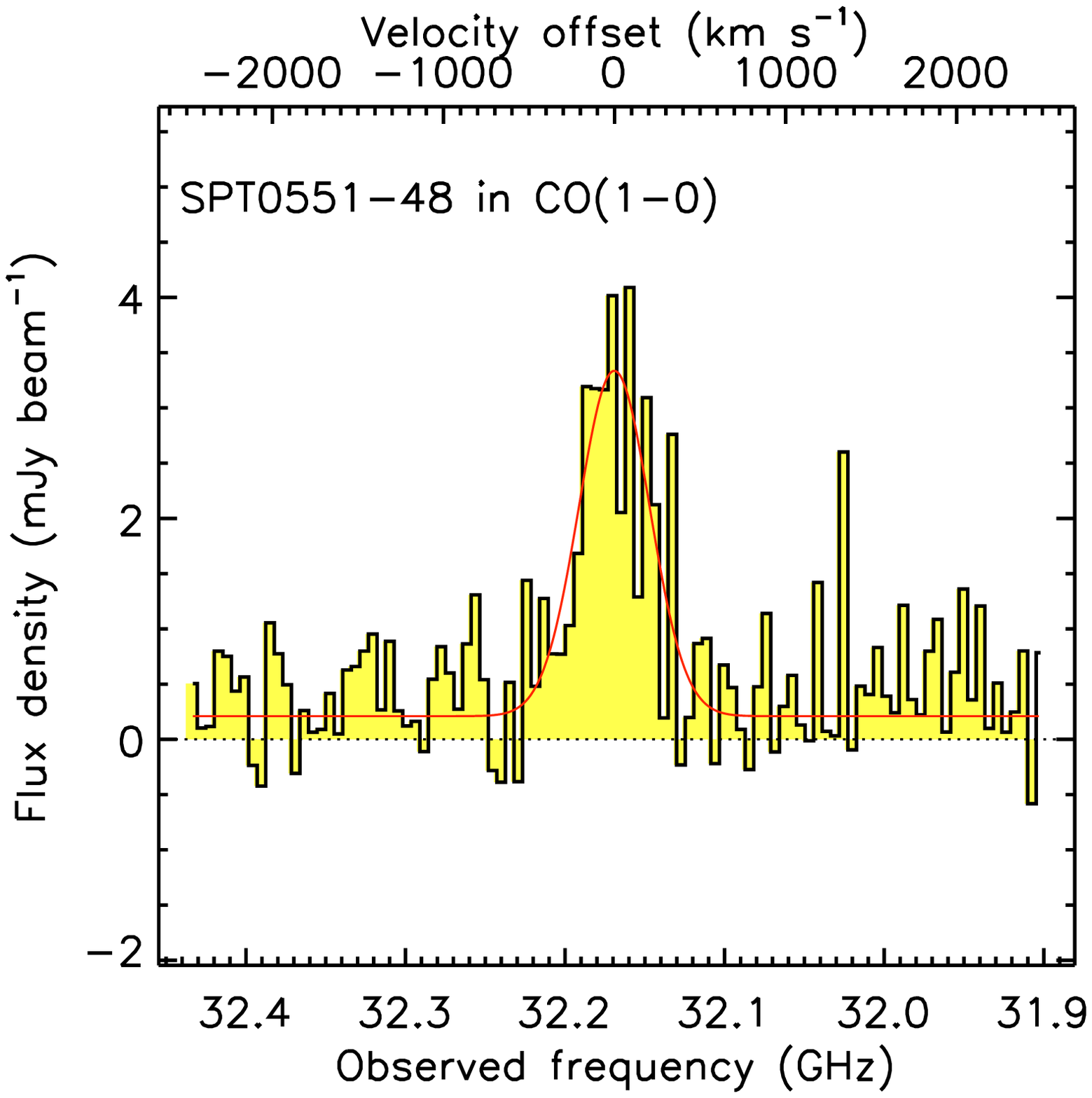}
\includegraphics[scale=0.33]{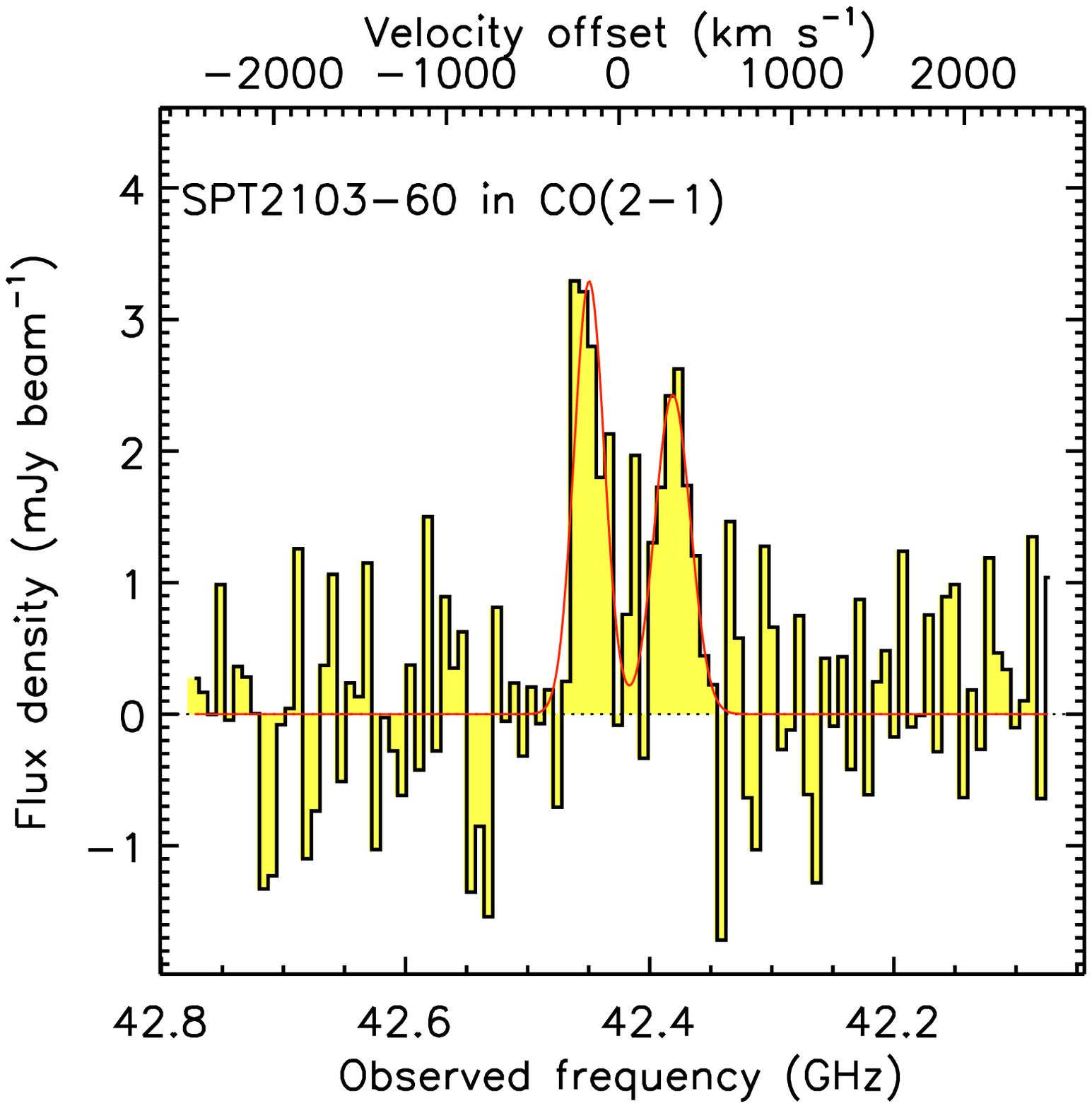}
\includegraphics[scale=0.33]{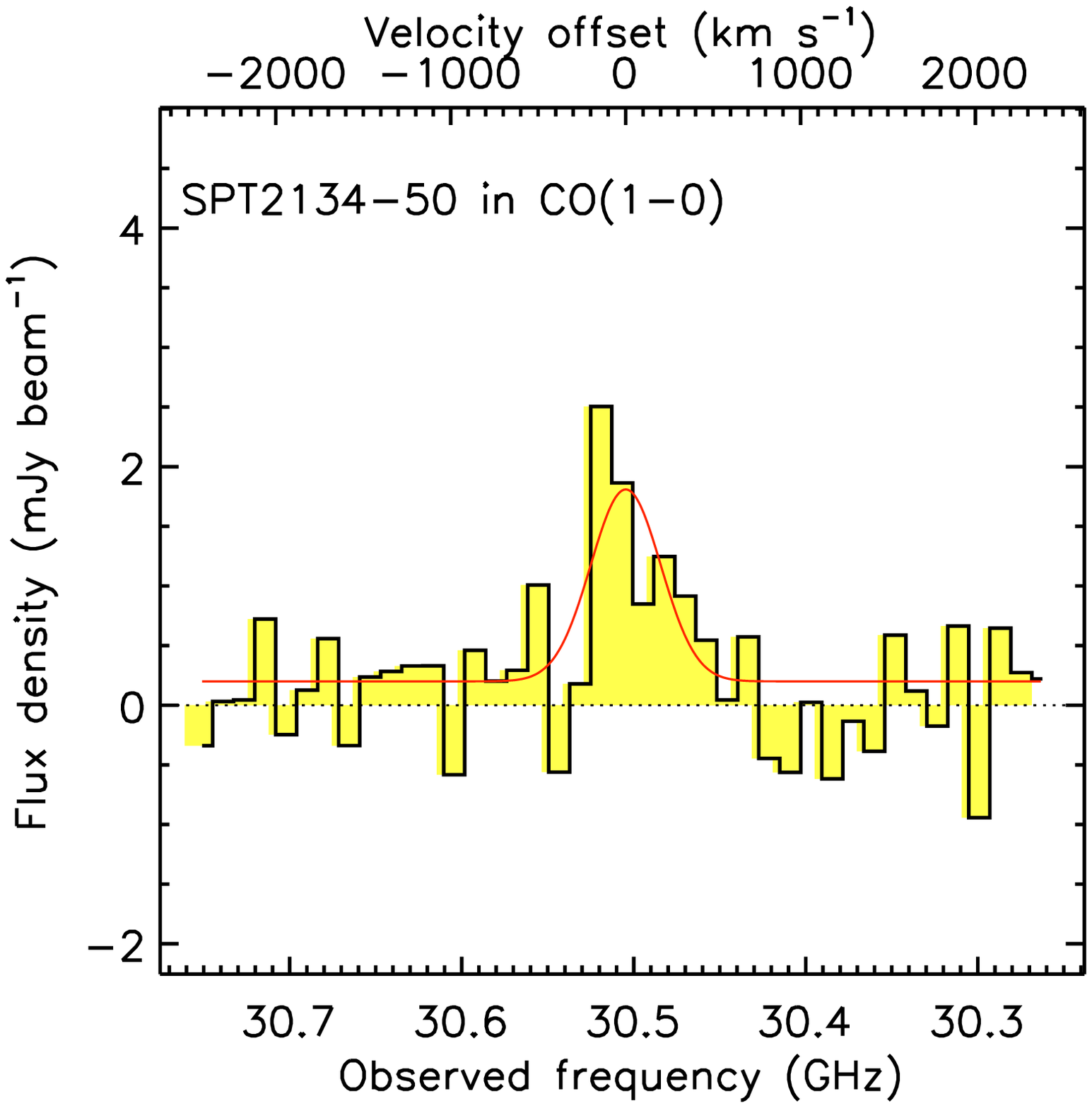}
\vspace{3mm}
\caption{CO (1--0/2--1) emission line spectra obtained with ATCA in our target SPT SMGs. All spectra are shown in 50 km s$^{-1}$ channels, except for SPT2134-50, which is shown at 120 km s$^{-1}$ channel resolution. Single or double Gaussian fits are represented by the red lines. The reference for $v=0$ km s$^{-1}$ in the upper x axis has been obtained from the inferred ATCA CO redshift.}
\end{figure*}

\begin{figure*}
\centering
\vspace{3mm}
\hspace{3mm}
\includegraphics[scale=0.33]{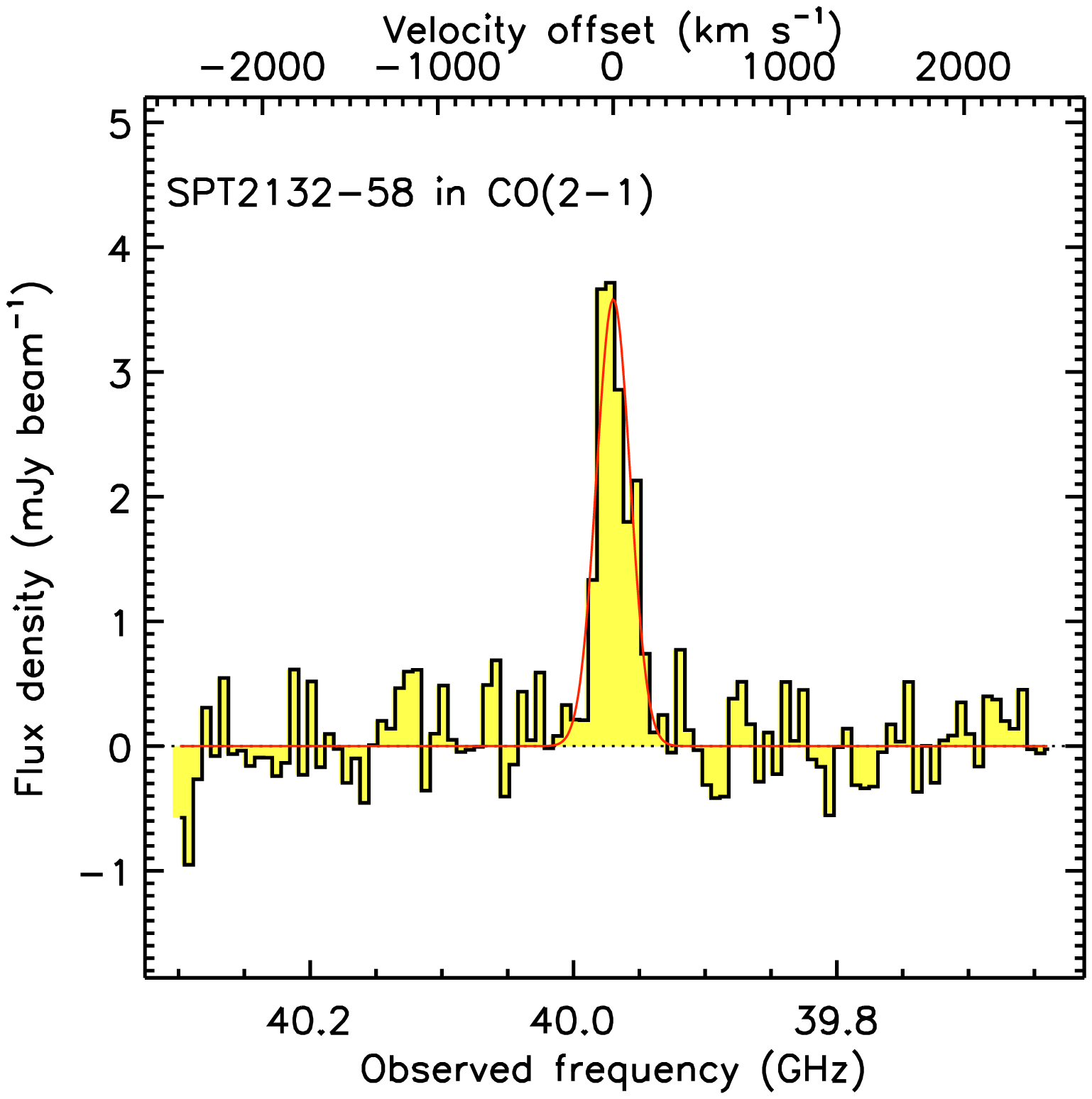}
\includegraphics[scale=0.33]{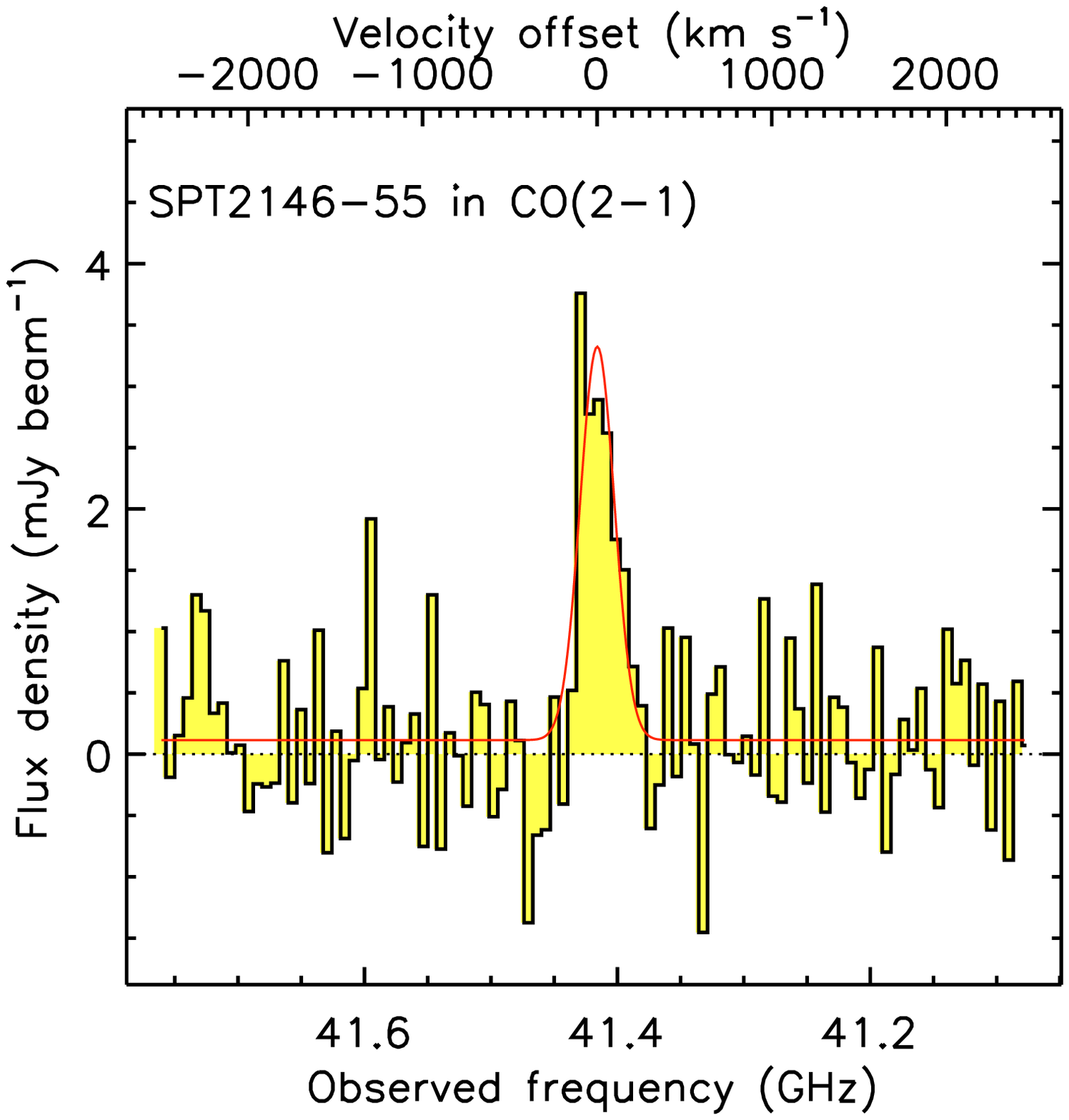}
\includegraphics[scale=0.33]{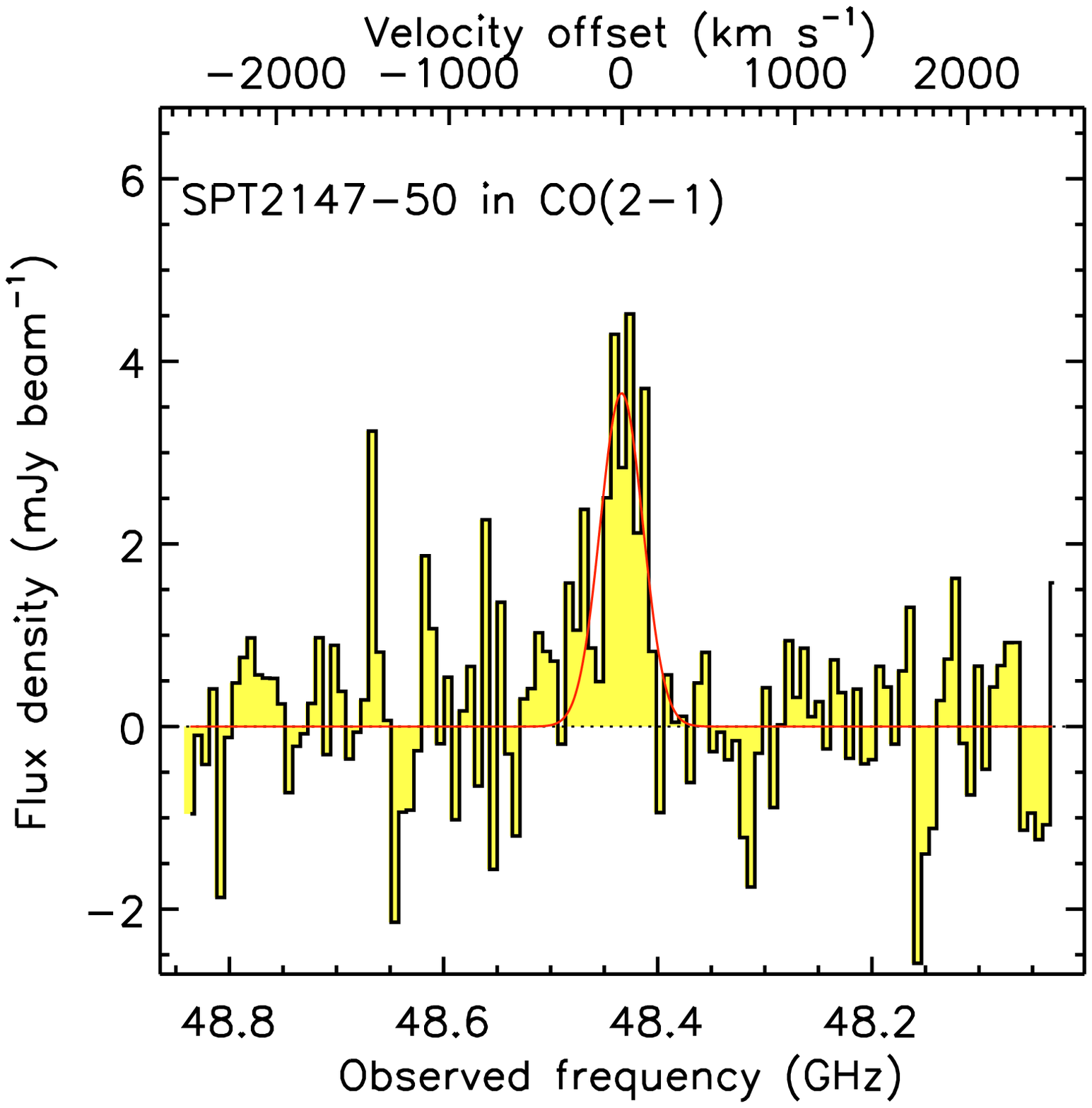}
\vspace{3mm}
\contcaption{}
\end{figure*}

\section{Observations and data}
\label{sect:data}
\subsection{Sample sources}

Our sample is drawn from the brightest sources discovered at 1.4-mm wavelength in the multi-band SPT Sunyaev-Zeldovich (SZ) survey of the southern sky, for which the spectral indexes from the mm photometry were consistent with DSFGs at high-redshift \citep[for details see][]{vieira10,mocanu13}. For a subset of 26 of these sources, follow-up spectroscopic observations covering the full 3-mm band ($\sim85-115$ GHz) were obtained with ALMA in order to derive redshifts based on the CO and [CI] line emission \citep{weiss13}. For 18/26 sources, it was possible to derive redshifts based on the identification of two or more emission lines. For 5/26 other sources, only one line was identified, and thus the redshifts were constrained using FIR/submm based photometric redshifts \citep[see][]{weiss13}. In the remaining 3/26 sources, no lines were identified \citep{weiss13}. 

For 3 other sources, redshifts were identified using spectroscopic observations covering the full 1-mm band ($\sim190-310$ GHz) with the Z-spec instrument on the Atacama Pathfinder Experiment (APEX) 12m telescope, and confirmed with optical spectroscopy obtained with the Very Large Telescope. These 3 sources share the same selection criteria as the ALMA sample of 26 sources \citep[see ][]{weiss13}. Observations of the CO(1--0) emission in 2 of these 3 objects, SPT-S\,233227$-$5358.5 and SPT-S\,053816$-$5030.8, were studied in detail in \citet{aravena13}. This makes a total of 18+3=21 SPT sources with confirmed, unambiguous redshifts obtained from millimeter spectroscopy.

Our survey sample is composed of 17 of the 21 SPT sources with redshifts spectroscopically confirmed from ALMA and APEX/Z-spec observations. In these cases, the sources redshifts allowed us to directly target the low-J CO emission lines with the ATCA 7-mm receivers (30-50 GHz). The remaining 4 sources lie at redshifts where neither low-J CO lines could be observed ($z=2.84-3.61$). One of these 4 sources correspond to SPT0551-50, which was previously identified at $z=2.1$ but has recently been revised as $z=3.1$ (Strandet et al., in prep.) 

A wealth of multi-wavelength data have been obtained for this sample of galaxies. These include optical spectroscopy of the lens galaxies with the VLT, Gemini and Magellan telescopes and optical/near-infrared imaging with the Hubble Space Telescope (HST); mid-infrared imaging with the {\it Spitzer} Infrared Array Camera (IRAC) at 3.6 and 4.5 $\mu$m; far-infrared imaging with the Photodetector Array Camera and Spectrometer (PACS) at 100 and 160 $\mu$m, and with the Spectral and Photometric Imaging Receiver (SPIRE) at 250, 350 and 500 $\mu$m on board of the {\it Herschel} space observatory. Millimeter and submillimeter imaging were obtained with the SPT at 3, 2 and 1.4-mm, and with the Submillimeter APEX Bolometer Camera (SABOCA) and the Large Bolometer Camera (LABOCA) at 350 and 870$\mu$m. For these, independent flux estimates were also obtained with ALMA at 870$\mu$m and 3-mm \citep{weiss13}.

\subsection{ATCA CO data}

We used ATCA in its H214 hybrid array configuration to observe either the CO(1--0) or the CO(2--1) emission line ($\nu_{\rm rest} = 115.2712$ and 230.5380 GHz, respectively) in the galaxies in our sample. We used the ATCA 7-mm receivers, which can be tuned in the frequency range 30--50 GHz. This frequency range covers the redshift ranges $1.38-2.84$ for CO(1--0) and $3.61-6.68$ for CO(2--1). Sources with redshifts in the range $z=2.84-3.61$ could not be observed in either of these low-J CO lines. The H214 array configuration at these observing frequencies leads to typical beam sizes of 5-6'', and thus preclude spatially resolving our sources. A summary of our observations is shown in Table 1.

The observations were performed as part of projects ID C2744 and C2818 during the periods 1-10 October 2012, 21-31 March 2013  and 1-7 April 2013, and were taken under mostly good weather conditions (atmospheric seeing values $90-400\ \mu$m) with 5 working antennas. 

We used the Compact Array Broadband Backend (CABB) configured in the wide bandwidth mode \citep{wilson11}. This leads to a total bandwidth of 2 GHz per correlator window and a spectral resolution of 1 MHz per channel ($\sim6-10$ km s$^{-1}$ per channel for the relevant frequency range). In all sources, one of the windows was tuned to observe the CO line, while the other window was tuned to measure the continuum emission or to target another fainter molecular line when possible. Individual tuning frequencies were estimated using the CO-based redshifts from ALMA and APEX/Z-spec spectroscopy \citep{weiss13}. The target positions were obtained from the ALMA 3-mm continuum detections \citep{weiss13}. 

Gain and pointing calibration were performed every 7--10 min and $\sim1$ hr, respectively. The gain/pointing calibrators used are listed in Table 1. The bright sources 0537-441 and 1921-293 were used as bandpass calibrators, and Uranus and 1934-638 were used as amplitude calibrators. We expect the flux calibration to be accurate to within 15\%, based on the comparison of the Uranus and 1934-638 fluxes. The software package {\it Miriad} \citep{sault95} and the Common Astronomy Software Applications \citep[][]{mcmullin07} were used for editing, calibration and imaging. 

The calibrated visibilities were inverted using the CASA task \verb CLEAN \, using natural weighting and cleaning in a tight box around the source position, down to a threshold of $\sim2.0\sigma$, where $\sigma$ is the rms noise level. The final rms and synthesized beam sizes are listed in Table 1.

\subsection{Lensing models}

ALMA 870$\mu$m continuum imaging at 0.5'' resolution were used to construct gravitational lens models of these objects of all the DSFGs in this study (Spilker et al, in prep).  Lens modelling was performed in the measured visibilities, allowing us to account for residual calibration uncertainties inherent in interferometric measurements and avoiding biases introduced by correlated noise when inverting visibilities into the image plane. The procedure used to derive the lens models is similar to the one used by \citet{hezaveh13}. All sources were significantly detected in the ALMA images (SNR $>50$), providing excellent constraints to these models. Optical and near-infrared spectroscopy of the foreground lenses will be presented in Rotermund et al. (in prep). The derived magnifications derived are listed in table \ref{table:properties3}.

The ALMA data presented in Spilker et al. (in prep) have much higher spatial resolution (0.5'')  and sensitivity compared to that presented by \citet{hezaveh13}, which used only the data taken in the compact configuration at a resolution of 1.5''. As such, the lens models from Spilker et al. are of much better quality. This explains the discrepancy between the magnification for SPT0418-47 of $\sim21$ derived by Hezaveh et al., and the new value of $\sim32$ presented in Spilker et al., in prep (see Table \ref{table:properties3}). Note that in the other two sources with lens models in Hezaveh et al. covered here (SPT0346-52 and SPT0538-50) the derived magnifications are fully consistent.

\subsection{IR data and models}

The far-infrared and (sub-)millimeter data were used to obtain spectral energy distribution (SED) models of the dust emission in the SPT DSFGs \citep[see][]{greve12, gullberg15}. The infrared luminosities ($L_{\rm IR}$), dust temperatures ($T_{\rm d}$) and masses ($M_{\rm d}$) used in this study were derived by fitting a single-component modified black-body dust model to the data (Strandet et al., in prep). We adopt a dust absorption coefficient $\kappa=0.045\times(\nu_{\rm r}/250 {\rm GHz})^\beta$ in units of m$^2$ kg$^{-1}$, with $\nu_{\rm r}$ the rest-frame frequency in GHz. We use a fixed emissivity index $\beta=2.0$, with the opacity equals to unity at $\lambda_0=100$ $\mu$m . Following \citet{greve12}, we fit the data only to $\lambda_{\rm rest}>50$ $\mu$m. Infrared luminosities are obtained by integrating the modelled SED in the range 8-1000\ $\mu$m rest-frame. Our SPT DSFGs have complete coverage from $\lambda_{\rm obs}=$ 250 to 3000 $\mu$m \citep{weiss13} allowing us for an accurate determination of the dust properties based on seven photometric data points.

To compare with similar high-redshift galaxy samples, we compiled far-infrared photometry for lensed DSFGs and main-sequence galaxies with published low-J CO detections \citep{ivison10,magdis12, magnelli12,harris12,combes12,johansson12,fu12,riechers13,rawle14}, and derived $T_{\rm d},\ M_{\rm d}$ and $L_{\rm IR}$ values using the same procedure as that applied to our SPT sources. These sources were compiled from the various deep extragalactic {\it Herschel} surveys, and have accurate and complete photometry from {\it Herschel} and other (sub)millimetre facilities \citep[see also][]{gullberg15}.

\begin{table*}
\centering
\caption{Observed line and continuum properties\label{table:properties}}
\begin{tabular}{lccccccc}
\hline
Source           &  Transition & $z_\mathrm{CO}$ $^a$ & $v_\mathrm{FWHM}$ $^b$ & $I_{\mathrm{CO}}$ $^c$ & $L_\mathrm{CO}'$ $^d$            & $S_{\nu}$ $^e$ & $\nu_{\rm cont}$\\
Short name                        &                     &                                        & (km s$^{-1}$)                       & (Jy km s$^{-1}$)                    & ($l_0$)  & ($\mu$Jy) & (GHz) \\
        \hline\hline
SPT0113-46  $^\ddagger$ & 2--1 & $4.2334 (3)$  &   $390\pm42$               &   $1.70\pm0.13$                    &   	$2.63\pm0.20$		&  $125\pm20$ & 43.0 \\
SPT0125-47  & 1--0 & $2.5146 (1)$  &   $428\pm27$               &   $2.70\pm0.22$                    &   	$7.93\pm0.65$		&  $230\pm25$ & 35.2\\
SPT0243-49  & 2--1 & $5.6965 (5)$  &   $598\pm47$               &   $1.00\pm0.08$                    &  	$2.43\pm0.19$		&  $120\pm10$ & 37.1\\
SPT0345-47  & 2--1 & $4.2958 (4)$ &    $357\pm55$               &   $1.80\pm0.20$                    &  	$2.85\pm0.32$		&  $170\pm35$ & 42.5\\
SPT0346-52  & 2--1 & $5.6551 (3)$ &    $613\pm30$               &   $2.15\pm0.15$                    & 	$5.16\pm0.36$		&  $160\pm20$ & 37.1\\
SPT0418-47  & 2--1 & $4.2253 (1)$ &    $324\pm19$               &   $1.30\pm0.12$                    &   	$2.00\pm0.18$		&  $145\pm20$ & 43.0\\
SPT0441-46  & 2--1 & $4.4812 (6)$ &   $552\pm77$               &   $0.95\pm0.14$                    &   	$1.60\pm0.24$		&   $130\pm25$ & 43.0\\
SPT0452-50  & 1--0 & $2.0078 (3)$ &    $612\pm59$               &   $0.96\pm0.12$                    &   	$1.91\pm0.24$		&  $<50$ & 37.4\\
SPT0459-59  & 2--1 & $4.7995 (4)$ &     $561\pm43$               &   $1.10\pm0.08$                    &    $2.06\pm0.32$	         &  $55\pm15$ & 38.7\\
SPT0538-50$^\dagger$ $^\ddagger$  & 1--0   & $2.7855(1)$ & $350\pm50$      &  $1.20\pm0.20$                       &    $4.18\pm0.70$          &  $140\pm20$ & 32.8\\
SPT0551-48  & 1--0 &$2.5833 (2)$  &   $485\pm40$                &   $1.40\pm0.20$                    &   	$4.29\pm0.61$	         & $200\pm50$ & 31.8\\
SPT2103-60  $^\ddagger$& 2--1 & $4.4340 (3)$  &    $476\pm37$          &   $1.60\pm0.25$                    &   	$2.66\pm0.41$		       &  $<135$ & 41.9\\
SPT2132-58  & 2--1 & $4.7678 (2)$ &       $225\pm17$         &   $0.85\pm0.07$                   &   		$1.58\pm0.13$	               &  $<70$ & 39.3\\
SPT2134-50  & 1--0 & $2.7788 (6)$ &      $469\pm180$        &   $1.00\pm0.18$                    &   	$3.48\pm0.63$		       &  $140\pm20$ & 31.9\\
SPT2146-55  & 2--1 &$4.5664 (3)$ &           $231\pm37$	     &   $0.95\pm0.16$                    &   	$1.65\pm0.28$		       &  $<120$ & 41.9\\
SPT2147-50  & 2--1 & $3.7599 (4)$ &     $290\pm52$           &   $1.25\pm0.25$                    &   		$1.60\pm0.32$	               &  $140\pm40$ & 47.5\\
SPT2332-53$^\dagger$  & 1--0   & $2.7256(2)$     & $342\pm42$ 	  &  $1.70\pm0.25$      &      $5.77\pm0.85$                & $<100$ & 33.2\\
\hline
\hline
\end{tabular}\\
\begin{flushleft}
\noindent $^a$ Redshift from the low-J CO lines. \\
\noindent $^b$ CO line full-width half maximum (FWHM) velocity. \\
\noindent $^c$ Integrated CO line intensity ($I_\mathrm{CO}=\int S_\mathrm{CO} dv$).\\
\noindent $^d$  Observed CO line luminosity in units of $l_0$=($\times10^{11}$ (K km s$^{-1}$ pc$^2$)$^{-1}$\\
\noindent $^e$ Continuum flux at the observed frequency $\nu_{\rm cont}$. \\
\noindent $^\dagger$ Observed properties taken from \citet{aravena13}. \\
\noindent $^\ddagger$ Velocity difference between the two line peaks is given instead of CO FWHM. \\
\end{flushleft}
\end{table*}

\section{Results}
\label{sect:results}
The obtained CO spectra are shown in Figure 1. Total integrated maps were obtained by collapsing the data cube along the frequency axis in the relevant range containing the detected CO emission line. Total intensities were then computed by spatially fitting a 2-dimensional Gaussian to these integrated maps using the task JMFIT from the Astronomical Image Processing System \citep[AIPS; ][]{greisen90}. Since all sources are unresolved at the achieved resolution, the CO spectra shown in Fig. 1 are obtained by measuring the fluxes at the CO peak position of the integrated images. To obtain the CO line central frequency and width, we fit single and double component Gaussian curves to the CO spectrum profiles. 

The CO spectra of most sources are consistent with single peaked Gaussian profiles. However, there are two sources where the profile and significance of the detection is consistent with double peaked profiles, SPT0113-46 and SPT2103-60. Two other sources, SPT0452-50 and SPT0441-46,  also show tentative signs of double-peaked CO profiles, but the evidence for multiple components is small at the depth of these data.

The spectrum of SPT2134-50 is shown averaged over broader channels in Fig.1. In this case, the significance of the detection is $\sim5\sigma$. The CO(1--0) line profile was checked against that measured in the CO(3--2) line from the ALMA 3mm spectroscopy \citep{weiss13}; we confirm that the line width and strength of the lines are consistent. In particular, the ratio between both CO lines, $R_{31}=L'_{\rm CO 3-2}/L'_{\rm CO 1-0}\sim1$ is comparable to that observed in other star-forming galaxies at high-redshift \citep[e.g.,][]{bothwell13}. 

We computed CO luminosities in units of K km s$^{-1}$ pc$^2$ using equation 3 from \citet{solomon97}. Note that $L'$ line luminosities are related to $L$ luminosities, in units of $L_\odot$, by $L/L' = (8\pi k_{\rm B}/c^2) \nu_{\rm rest}^3$, where $k_{\rm B}$ is the Boltzmann constant, $c$ is the speed of light and $\nu_{\rm rest}$ is the line rest-frame frequency. For the CO(1--0) line, this yields $L_{\rm CO}/L'_{\rm CO}=4.9\times10^{-5}$, with $L_{\rm CO}$ in units of $L_\odot$. For sources that have been detected in the $J=2-1$ transition, we convert to the ground level by assuming a line brightness temperature ratio between the CO(2--1) and CO(1--0) lines of 0.9. This ratio is consistent with previous findings for DSFGs where an average ratio of 0.85 was found for a sample of 32 sources \citep{bothwell13}, consistent within the uncertainties with the line ratio of $\sim0.7$ found for star forming disk galaxies at $z=1.5-2.0$ \citep{aravena10,aravena14}, and follows the average ratio of  $\sim1.1$ found through stacking of the SPT DSFGs themselves \citep{spilker14}. The measured intensities, luminosities, line-widths and frequencies are given in Table 2. 

Measurements of the continuum emission were obtained by collapsing the data along the frequency axis in the line-free SPW and channels around the line. Flux measurements were obtained by fitting a 2-dimensional Gaussian to the resulting continuum map. The derived continuum fluxes are listed in Table 2.

\section{Analysis}
\label{sect:analysis}
\subsection{Line widths}

\begin{figure}
\centering
\includegraphics[scale=0.55]{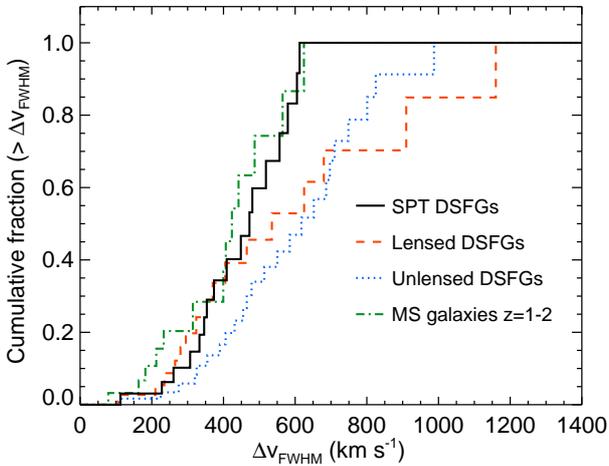}
\caption{Cumulative distribution of linewidths (FWHM) derived from the CO profiles for the SPT DSFGs, shown as a black solid histogram, compared with the cumulative distribution of linewidths for other high-redshift galaxy samples with available $J\leq2$ CO measurements. The orange dashed histogram shows the distribution for the lensed DSFGs from literature (see text). The blue dotted histogram shows a literature compilation of unlensed DSFGs and the green dot-dashed histogram represents a sample of MS star-forming galaxies at $z=1-2$. \label{fig:fwhm}}
\end{figure}

Previous studies have found a wide range of CO full-width at half maximum (FWHM) linewidths for DSFGs \citep{neri03, greve05, tacconi06, tacconi08}. The mean value for CO FWHM found by \citet{bothwell13} in the largest sample of (unlensed) DSFGs studied to date is $510\pm80$ km s$^{-1}$. Individual values range from 200--1400 km s$^{-1}$. Even though this study used mostly $J>2$ transitions of CO, a subsample of these galaxies were detected in $J=2$. For these sources, the mean CO FWHM is consistent with the full sample within the statistical uncertainties. 

For a more straightforward comparison, in Fig. \ref{fig:fwhm} we show the cumulative distribution of CO FWHM for different samples of galaxies that have been detected in low-J CO ($J\leq2$), including the comparison sample of lensed DSFGs described in Section 2.3  \citep{ivison10,harris12,combes12,rawle14,johansson12,fu12,riechers13}, a compilation of 23 unlensed DSFGs from the literature \citep{coppin10, carilli11,ivison11, ivison13,riechers11a, riechers11b,thomson12, bothwell13, hodge13}, and a sample of 13 main-sequence galaxies at $z=1.0-1.5$ \citep{daddi10,magnelli12}. 

Qualitatively, the distributions of all samples appear to be different, particularly when comparing with the unlensed DSFG sample. SPT DSFGs have a distribution concentrated at smaller linewidths, whereas most of the unlensed DSFGs from the literature show larger linewidths. Literature lensed DSFGs appear to follow the distribution of SPT DSFGs at linewidths $<400$ km s$^{-1}$, but then they depart to large linewidths. The SPT DSFGs show a weighted-average FWHM of $370\pm130$ km s$^{-1}$, where the uncertainty includes the scatter in the sample. This compares well with the literature lensed DSFGs, which show a weighted average FWHM of $320\pm290$ km s$^{-1}$, and is also compatible with the average for MS galaxies at $z<2$ of $320\pm160$ km s$^{-1}$. Unlensed DSFGs, however, show a larger average linewidth of $470\pm220$ km s$^{-1}$. This difference is obvious from Fig. \ref{fig:fwhm}.

To quantify the difference between the line width distribution of the various samples, we performed a Kolmogorov-Smirnov (KS) test. We measure the KS probability $P$ that a pair of datasets are drawn from the same distribution. Low values of $P$ indicate that both datasets are significantly different. We compared the FWHM distribution of the SPT DSFG sample with the literature lensed DSFGs, the MS galaxies and the unlensed DSFGs, and found $P=0.86, 0.60$ and 0.25, respectively. Such values indicate that the line width distribution of the SPT DSFGs is compatible with that of other galaxy populations, in particular, with literature lensed DSFGs (as expected) and MS galaxies. Furthermore, with $P=0.25$, there is suggestive evidence that the SPT and unlensed DSFG samples do not come from the same linewidth distribution. 

\subsection{$L'_{\rm CO}$ vs FWHM}

\label{sect:lens}
\begin{figure}
\centering
\includegraphics[scale=0.45]{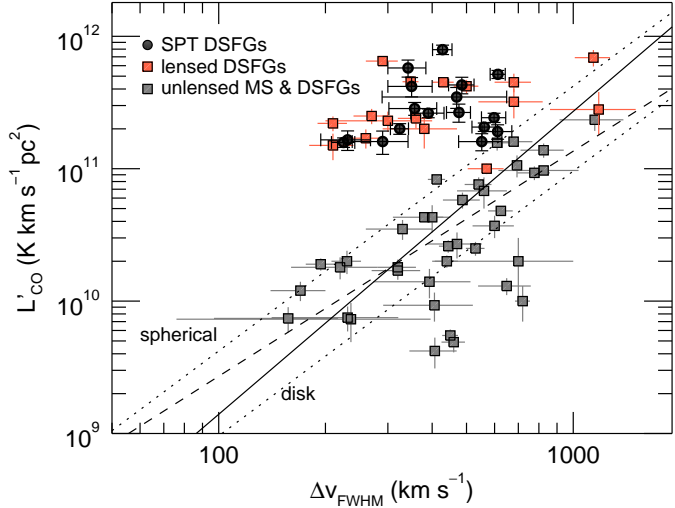}
\caption{$L'_{\rm CO}$ vs FWHM for our SPT DSFGs (black filled circles), compared to a literature compilation of lensed DSFGS (orange squares) and unlensed DSFGs (gray squares). The solid line shows the best fit relation obtained for unlensed DSFGs, which include in this case both SMGs and MS galaxies. The dashed line shows the best fit relation found by \citet{harris12} for unlensed DSFGs. The dotted lines show a simple ``virial'' functional form for the CO luminosity for a compact starburst and an extended disk (see text).  \label{fig:lco_fwhm}}

\end{figure}

\begin{figure*}
\centering
\includegraphics[scale=0.42]{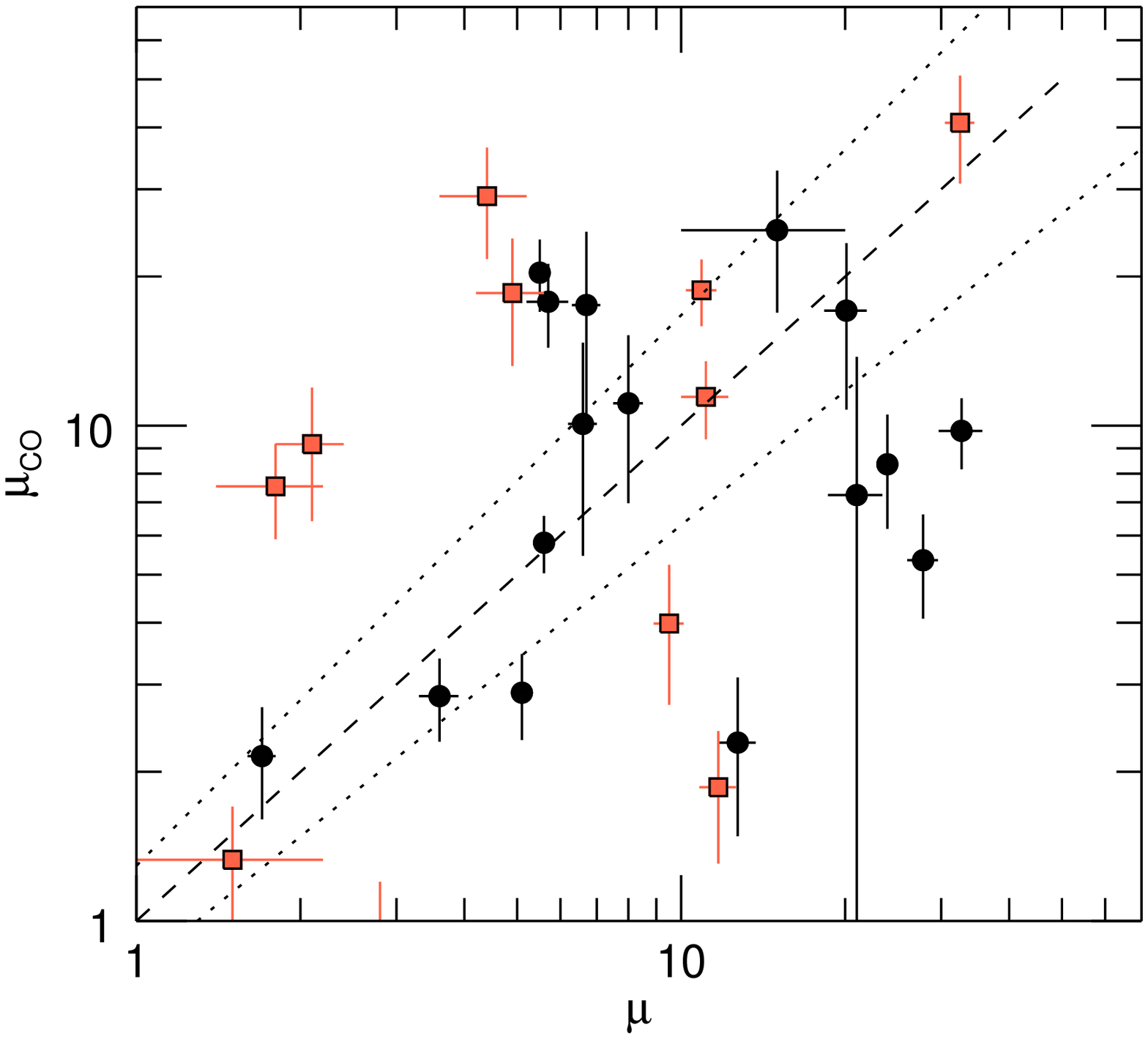}
\hspace{3mm}
\includegraphics[scale=0.42]{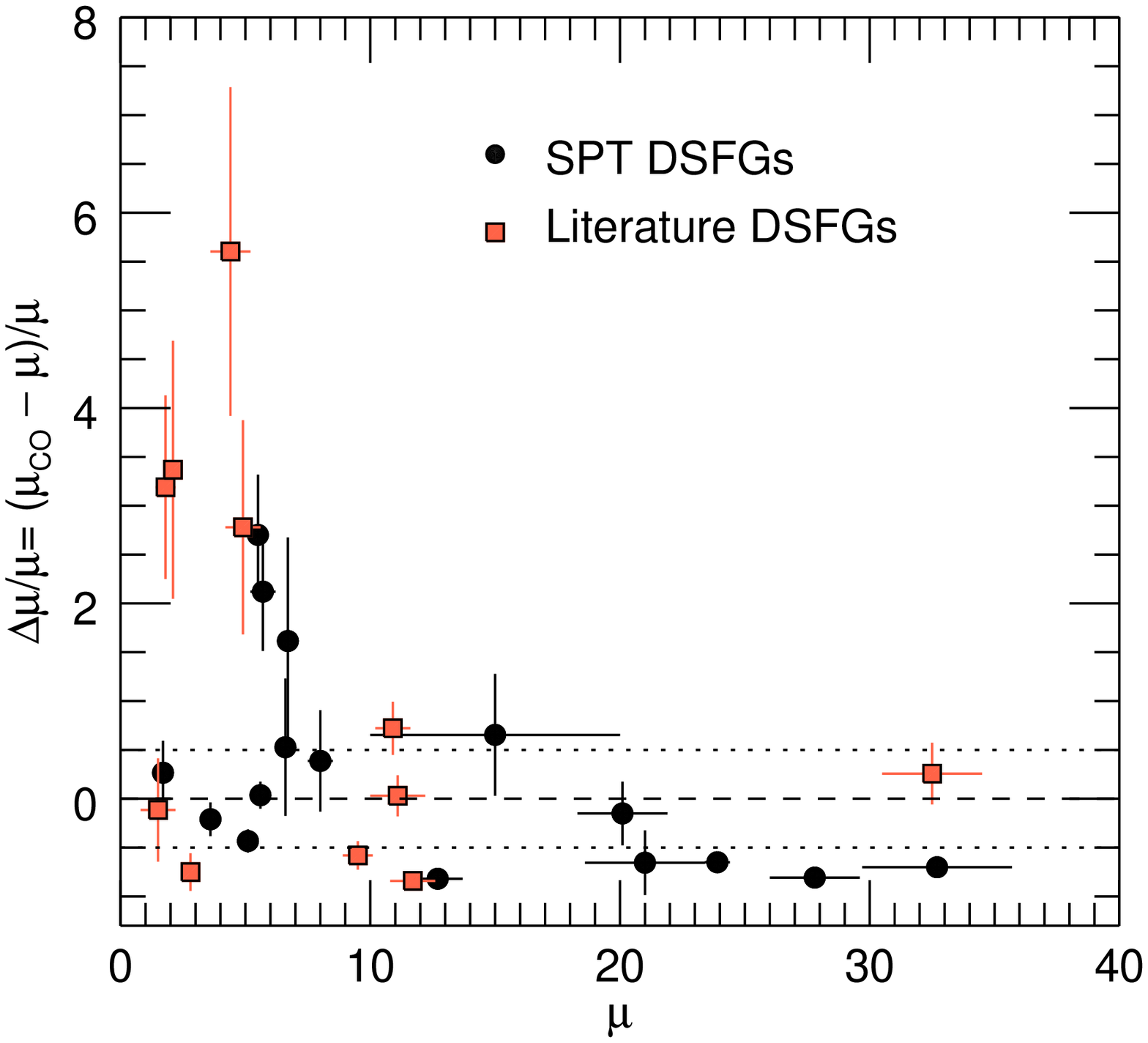}
\caption{Lensing magnification factors obtained by assuming the best fit $L'_{\rm CO}$ vs FWHM line, $\mu_{\rm CO}$, compared to the actual magnification obtained from lens modelling for SPT DSFGs (Spilker et al, in prep.). Also shown are lensed DSFGs from the literature for which a value of $\mu$ was available \citep{johansson12,fu12,bussmann13, riechers13}. ({\it Left:}) $\mu$ vs. $\mu_{\rm CO}$. ({\it Right:}) Fractional difference between $\mu$ and $\mu_{\rm CO}$. The dashed line represents the case where $\mu_{\rm CO}= \mu$. The dotted lines enclose cases where there is an agreement between both estimates of $<50\%$.}
\label{fig:mu_mu}
\end{figure*}

Based on physical arguments about the ability of the CO line emission to trace mass and kinematics, \citet{bothwell13} suggested an empirical relationship between the CO luminosity ($L'_{\rm CO}$) and the CO line width at FWHM ($\Delta v_{\rm FWHM}$), with the form $L'=a (\Delta v_{\rm FWHM})^b$, where a and b are parameters to be obtained from observational data..

Under the premise that this relationship holds for gravitationally lensed sources, \citet{harris12} suggested the use of this relationship to obtain a measurement of the magnification $\mu$, assuming the intrinsic (true) luminosity is related to the observed luminosity by 

\begin{equation}
L'_{\rm obs}=\mu L'_{\rm intrinsic}=\mu a (\Delta v_{\rm FWHM})^b.
\end{equation}

\noindent Therefore, $\mu=L'_{\rm obs}/a(\Delta v_{\rm FWHM})^b$ can be derived from CO measurements.

In this section, we test the method to obtain lens magnifications from CO, by comparing the true magnification obtained from lens modelling of the dust emission, $\mu_{870\mu\rm m}$, for lensed DSFGs with the magnification factors inferred from the CO linewidth-$L'$ relation, $\mu_{\rm CO}$.

Figure \ref{fig:lco_fwhm} shows the relationship between $L'_{\rm CO}$ and the line FWHM that is obtained for unlensed DSFGs and MS galaxies that have low-J CO detections \citep[][]{riechers11, riechers14, ivison13, ivison11, frayer08, thomson12, carilli11, hodge13, bothwell13, walter12, coppin10}. We specifically consider only sources with low-J CO measurements (CO 1--0 or CO 2--1) to avoid uncertainties from different line transitions having different linewidths. The solid line shows a best fit relationship to unlensed DSFGs and MS galaxies at $z=1-2$. The dashed line shows the best fit relationship presented in equation 2 of \citet[][]{harris12}, which includes unlensed DSFGs but not MS galaxies. Due to the significant scatter in individual CO luminosities that can span nearby an order of magnitude and uncertainties in line widths, both relationships provide a similar fit to the data points.

Following \citet{bothwell13}, we also show a simple parametrisation for $L'_{\rm CO}$:

\begin{equation}
L'_{\rm CO} =   C (\frac{\Delta v_{\rm FWHM}}{2.35})^2 \frac{R}{\alpha_{\rm CO} G},
\end{equation}

\noindent where $\Delta v_{\rm FWHM}$ is the CO line width at FWHM in km s$^{-1}$, $R$ is the radius of the CO emitting region in parsecs, $G$ is the gravitational constant, $\alpha_{\rm CO}$ is the CO luminosity to gas mass conversion factor in units of K km s$^{-1}$ pc$^2$ and $C$ is a constant that accounts for the geometry of the galaxy \citep{erb06,bothwell13}. In Fig. \ref{fig:lco_fwhm}, we overlay two cases that bracket the range of $L'_{\rm CO}$ for a given line FWHM: a disk galaxy model for which $C=2.1$, $R=5$ kpc, $\alpha_{\rm CO}=4.6$; and a virialized spherical source geometry with $C=5$, $R=2$ kpc, $\alpha_{\rm CO}=1.0$. 

Also shown in Fig. \ref{fig:lco_fwhm} are the CO measurements obtained for the SPT DSFGs and a compilation of gravitationally lensed DSFGs with low-J CO measurements \citep{harris12,ivison10,swinbank10,lestrade11,fu12}.  The most notable feature is the relatively flat distribution of $L'_{\rm CO}$ values with varying linewidths. This flat distribution suggests that galaxies with narrower line widths need larger magnification corrections. 

To obtain the magnification factor, one only needs to correct by the difference between the observed CO luminosity, and the unlensed $L'_{\rm CO}-$ FWHM relation for a given $\Delta v_{\rm FWHM}$. We used our best fit $L'_{\rm CO}-$FWHM relationship from Fig. \ref{fig:lco_fwhm} to compute a CO-based magnification estimate $\mu_{\rm CO}$ for the SPT DSFGs. The derived values are listed in Table \ref{table:properties3}. 

In Fig. \ref{fig:mu_mu} we quantify the accuracy of the $\mu_{\rm CO}$ values, by comparing them to the magnification values obtained through lens modelling of the dust continuum emission $\mu_{\rm d}$ where available \citep[][; Spilker et al., in prep]{swinbank10, lestrade11, fu12, hezaveh13, bussmann13}. 

The large scatter in the unlensed $L'_{\rm CO}-$FWHM relationship shown in Fig. \ref{fig:lco_fwhm} (about an order of magnitude) makes it highly unreliable to derive accurate magnification factors from CO measurements. For only 9 out of the 28 objects with available $\mu_{\rm d}$ values (32\% of the sample), the difference amounts to $\la$ 50\% (i.e. $\Delta \mu/\mu<0.5$ or $0.5\mu_{\rm d}<\mu_{\rm CO}<1.5\mu_{\rm d}$). This means that most sources have CO-derived magnification values that significantly differ from the values obtained from lens modeling. Interestingly, for several of the sources with $\mu_{\rm d}<8$ the CO-based estimate severely overpredicts the magnification ($\Delta \mu/\mu>1$). Even if we remove the outlier objects (with $\Delta \mu/\mu>1$), there will be 11 objects with $\Delta \mu/\mu>0.5$ and only 9 objects with $\Delta \mu/\mu<0.5$. 

The reason why the outlier sources show significant deviations with respect to $\mu_{\rm d}$ value is not clear, and there is no clear trend among the CO observables (e.g. $L'_{\rm CO}$, FWHM, signal to noise ratio) that could hint for an effective prior selection of outliers. We suggest two possible explanations for this exceedingly large discrepancies on the derived magnification factor: (1) inaccurate lensing models; and (2) unsampled parameter space in the $L_{\rm CO}$ versus FWHM plane for both lensed and unlensed sources.

First, this discrepancy might be related to inaccurate lensing models. Four of the sources with the larger $\mu$ deviations, which belong to the sample of \citet{harris12}, do have somewhat peculiar lens configurations as shown by recent high resolution submillimeter 880$\mu$m imaging \citep{bussmann13}. More specifically (see Bussmann et al. for details): J090302.9-014127 presents one of the most compact lens image separations of their sample, and it is possible that the lens itself contributes to the submillimeter emission; J091305.0-005343 corresponds to the largest background source in their sample; In J133649.9+291801, the optical images do not show a lens source, even though the submillimeter map shows evidence for strong lensing; and J141351.9-000026 shows no counter image in the submillimeter map being thus possibly not strongly lensed. Similarly, SPT2132-58 appears to be a compact source in the ALMA images and lens models (Spilker et al., in prep). Hence, it is possible that the largest differences between the lens-model and CO-based magnification estimates for low $\mu_{d}$ are due to a combination of effects, including an inaccurate lens-model for a complex lens configuration and/or due to differential lensing of more compact regions in the submillimeter continuum compared to those more extended seen in CO.

Secondly, the discrepancy in magnification factors could be related to a poorly sampled parameter space in the $L'_{\rm CO}-$FWHM plane for both lensed and unlensed sources. The measured $L'_{\rm CO}$ for lensed objects is limited to the depth of such observations. Deeper observations of currently CO undetected objects, might extend the range and scatter of $L'_{\rm CO}$ for lensed objects. Similarly, the current low-J CO observations of unlensed objects might not cover the full range of $L'_{\rm CO}$ versus FWHM, particularly for small line widths ($<500$ km s$^{-1}$) with significant CO luminosities (few $10^{10}$ K km s$^{-1}$ pc$^2$). In such cases, the current implementation of this $L'_{\rm CO}-$FWHM method may be flawed and thus would lead to large discrepancies for low $\mu_{\rm d}$.

In summary, our results show that even a rough estimation of the lensing magnification based on unresolved CO measurements is highly unreliable. The large scatter in the $L'_{\rm CO}$-FWHM relation (driven by the different source geometries), implies that such estimates will be typically off by factors of  $\sim2$, particularly when there is independent evidence that the lens configuration could be unusual. %In a statistical sense, 

\begin{table*}
\centering
\caption{Derived physical properties\label{table:properties3}}
\begin{tabular}{lccccccccc}
\hline
Source $^a$    & $\mu_{870}$ $^b$& $\mu_{\rm CO}$ $^c$& $T_d$ $^d$& $M_d$ $^d$& $L_{\rm IR}$ $^d$&  $M_{\rm gas}/\mu$ $^e$& $M_{\rm dyn}$ $^f$ & $\alpha_{\rm CO}$ $^g$ & $\alpha_{\rm CO, lim}$ $^h$\\
Short name     &                   &                                    & (K) & ($10^9\ M_{\odot}$)& ($10^{13}\ L_\odot$) & ($10^{10}\ M_{\odot}$) & ($10^{10}\ M_{\odot}$) & $\alpha_0$ & $\alpha_0$ \\
\hline
SPT0113-46   & $23.9\pm0.5$  & $9\pm2$       &  $33\pm1$ &  $3.7\pm1.0$ & $3.0\pm0.5$   & $0.9\pm0.1$   &  $2.2\pm0.5$   &  $1.2\pm0.4$ & $7.0$ \\
SPT0125-47   & $5.5\pm0.1$   & $21\pm4$     &  $42\pm2$ & $5.6\pm1.3$ & $12.3\pm1.6$  & $11.5\pm1.0$   &  $\ldots$            & $0.7\pm0.2$ & $0.7$\\
SPT0243-49   & $5.1\pm0.2$   & $3\pm1$      &   $35\pm1$ & $3.4\pm1.1$ & $4.5\pm0.8$    & $4.2\pm0.4$   &  $9.8\pm1.6$   & $1.3\pm0.4$ & $3.8$ \\
SPT0345-47   & $8.0\pm0.5$   & $11\pm4$     &   $56\pm3$& $1.7\pm0.4$ & $13.0\pm2.4$  & $3.2\pm0.4$    &  $2.0\pm0.6$     & $0.5\pm0.2$ & $1.8$\\
SPT0346-52   & $5.6\pm0.1$   & $6\pm1$       &  $55\pm2$ & $2.0\pm0.6$ & $16.2\pm2.4$  & $8.2\pm0.6$    &  $6.3\pm0.6$  & $0.4\pm0.1$  & $2.1$ \\
SPT0418-47   & $32.7\pm3.0$ & $10\pm2$       &  $49\pm2$ & $2.5\pm0.7$ & $7.7\pm1.3$    & $0.5\pm0.1$    &   $1.9\pm0.3$   & $1.1\pm0.3$ & $8.7$\\
SPT0441-46   & $12.7\pm1.0$ & $2\pm1$       &  $41\pm2$ & $2.4\pm0.7$ & $4.8\pm0.9$    & $1.1\pm0.2$     &  $4.3\pm1.2$   & $1.3\pm0.4$  & $12.3$\\
SPT0452-50   & $1.7\pm0.1$   & $2\pm1$       &  $23\pm1$ & $3.4\pm0.8$ & $0.7\pm0.1$    & $9.0\pm1.2$    &    $23.4\pm4.7$  & $1.8\pm0.5$ & $1.9$ \\
SPT0459-59   & $3.6\pm0.3$   & $3\pm1$       &  $41\pm2$ & $1.8\pm0.5$ & $4.0\pm0.8$    & $5.1\pm0.6$    &    $21.9\pm3.9$  & $0.8\pm0.2$  & $2.8$ \\
SPT0538-50   & $20.1\pm1.8$ & $17\pm6$     &  $39\pm1$ & $5.1\pm1.2$ & $8.0\pm1.0$    & $1.7\pm0.3$    &    $4.5\pm1.3$    & $1.2\pm0.3$ & $3.3$\\
SPT0551-48   & $\ldots$          & $8\pm2$       &  $42\pm2$ & $5.4\pm1.2$ & $11.0\pm1.6$   & $3.4\pm1.8$    &    $\ldots$           & $1.3\pm0.3$  & $3.1$\\
SPT2103-60   & $27.8\pm1.8$  & $5\pm1$       &  $40\pm1$ & $2.5\pm0.7$ & $4.4\pm0.7$     & $0.9\pm0.1$    &    $3.1\pm0.5$     & $0.8\pm0.3$  & $12.0$\\
SPT2132-58   & $5.7\pm0.5$  & $18\pm4$     &  $41\pm2$ & $1.7\pm0.5$ & $4.2\pm0.7$     & $2.5\pm0.3$   &      $0.9\pm0.2$    &  $0.9\pm0.3$ & $0.9$\\
SPT2134-50   & $21.0\pm2.4$ & $7\pm6$       & $43\pm2$  & $3.7\pm0.9$ & $7.2\pm1.0$    & $1.3\pm0.3$   &      $1.6\pm1.2$   &  $1.1\pm0.3$  & $7.5$\\
SPT2146-55   & $6.7\pm0.4$   & $18\pm7$     &  $49\pm2$ & $1.7\pm0.5$ & $3.6\pm0.8$    & $2.2\pm0.4$  &       $1.3\pm0.4$   &  $0.9\pm0.3$ & $1.1$\\
SPT2147-50   & $6.6\pm0.4$   & $10\pm5$     & $42\pm2$  & $1.9\pm0.5$ & $4.1\pm0.6$    & $2.2\pm0.5$    &     $2.5\pm1.6$  & $1.1\pm0.3$  & $1.7$\\
SPT2332-53   & $\ldots$          & $25\pm8$     & $49\pm2$  & $5.3\pm1.2$ & $12.2\pm1.6$ & $4.6\pm2.4$   &      $\ldots$         & $0.9\pm0.3$  &  $1.1$\\
\hline
\end{tabular}\\
\begin{flushleft}
\noindent $^a$ Source name. \\
\noindent $^b$ Magnification factor obtained through visibility-based lens modelling of ALMA 870$\mu$m continuum imaging data at 0.5'' resolution.\citep[Spilker et al. in prep; ][]{hezaveh13}\\
\noindent $^c$ CO-derived magnification factor, computed using the following relation: $\mu_{\rm CO}=3\times(L'_{\rm CO}/10^{11})(\Delta v_{\rm FWHM}/400)^{-2.3}$, where $L'_{\rm CO}$ is the measured CO luminosity in units of (K km s$^{-1}$ pc$^2$) and $\Delta v_{\rm FWHM}$ is the measured linewidth at FWHM in km s$^{-1}$. See Section \ref{sect:lens}.\\
\noindent $^d$ Dust properties derived from multi-wavelength SED fitting. The parameters listed here have not been corrected for lensing magnification.\\
\noindent $^e$ Gas mass derived from the observed CO luminosity assuming $\alpha_{\rm CO}=0.8$, and divided by the lensing magnification $\mu$ where available. A typical value of $\mu=10\pm5$ was used when a lensing model was not available, for SPT0551-48 and SPT2332-53. \\
\noindent $^f$  Dynamical mass estimated from the line FWHM and the lensing derived effective radius (Spilker et al., in prep), and assuming a virialized spherical geometry as described in the text. No estimate is provided for SPT0125-47 given the complex source geometry seen in the source-plane ALMA 870$\mu$m image.\\
\noindent $^g$ Dust-derived CO luminosity to gas mass conversion factor, in units $\alpha_0=M_\odot$ (K km s$^{-1}$ pc$^2$).\\
\noindent $^h$ Limit to the CO luminosity to gas mass conversion factor based on a maximum dynamical mass.\\
\end{flushleft}
\end{table*}

\subsection{CO and dust sizes}
\label{sect:sizes}
Assuming that the CO line emission is optically thick, it is possible to estimate the CO emitting size, $r_{\rm CO}$, even if our observations are unresolved. From the derivation of equation 2 in \citet{solomon05}, and assuming the CO lines have single Gaussian profiles, we find $L'_{\rm CO}\approx1.13 \mu (T_{\rm ex}-T_{\rm CMB}) \Delta v_{\rm FWHM} r_{\rm CO}^2$, where $L'_{\rm CO}$ is the line luminosity in units of K km s$^{-1}$ pc$^2$, $\mu$ is the lensing magnification factor, $T_{\rm ex}$ is the gas excitation temperature, in Kelvin, that we assume to be mean dust temperature within the SPT sample, $\sim40$ K, $T_{\rm CMB}=2.73(1+z)$ is the Cosmic Microwave Background (CMB) temperature at redshift $z$, $\Delta v_{\rm FWHM}$ is the line width at FWHM measured in km s$^{-1}$ and $r_{\rm CO}$ is the size of the CO emitting region in parsecs. This expression allows us to estimate the magnification if there is a good CO size measurement. Note also that the expression for $r_{\rm CO}$ depends weakly on the assumed value of $T_{\rm ex}$, and thus a change of a factor $2$ of this parameter will yield a $\sqrt{2}$ variation on the CO size. 

\begin{figure}
\centering
\includegraphics[scale=0.42]{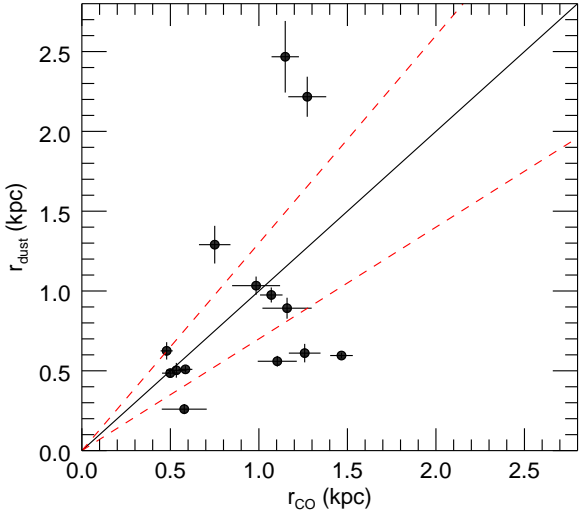}
\caption{Comparison of the CO and dust sizes for the SPT DSFGs (see Section \ref{sect:sizes}). The red dashed lines enclose the location where both estimates differ by less than 30\% (i.e. CO sizes are 0.7 and 1.3 times larger than the dust sizes).}\label{fig:sizes}
\end{figure}

In Fig. \ref{fig:sizes} we compare the CO size estimated using the above relation with the source sizes derived from the lens reconstruction of the dust emission using the ALMA  870$\mu$m continuum maps, where the latter were available (Spilker et al., in prep). The CO sizes appear to be up to $2\times$ larger for most sources; however, 4 objects appear to have CO sizes smaller than the dust sizes. This is similar to what is observed in the literature \citep[e.g. see detailed discussion by][]{spilker15}.  

A possible reason for these discrepancies is that galaxies can be morphologically different at different wavelengths and thus the lensing magnification factors for CO and dust could differ significantly, by up to $\sim50\%$ \citep{spilker15}.  Similarly, differences in the beam filling factors for CO and dust emission may lead to differences in the sizes derived, particularly for our CO size estimate, which does not resolve the source. Note also that a smaller filling factor at fixed flux will not change the CO size estimate, which is based on the flux measurement, and the assumption of $T_{\rm ex}=T_{\rm d}$ and optically thick CO), but will increase dust size estimate, which reflects an angular size that will increase as filling factor gets smaller. However, resolved CO imaging of the SPT DSFGs is needed to infer more quantitative conclusions, particularly since the CO size estimation assumes that the dust-derived magnification applies for CO  (i.e. assumes no differential lensing) and that the CO linewidth can be approximated as a single Gaussian. The ALMA dust continuum maps provide the more accurate information about the source sizes currently available. Hence, we do not use the CO sizes in the following sections and only use the ALMA-derived dust sizes throughout. 

\subsection{Dynamical masses}

\label{sect:mdyn}

\begin{figure*}
\centering
\includegraphics[scale=0.42]{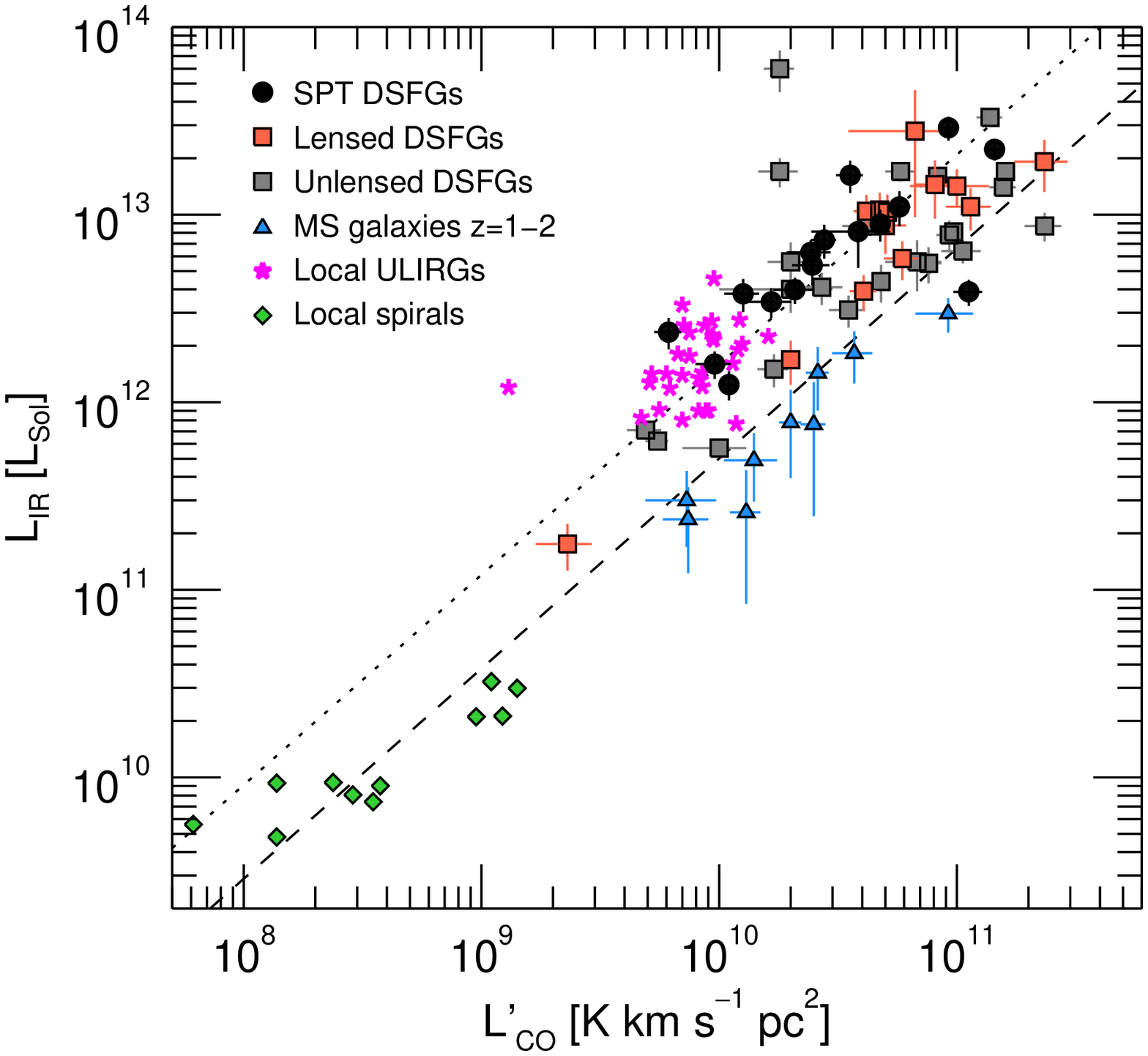}
\includegraphics[scale=0.42]{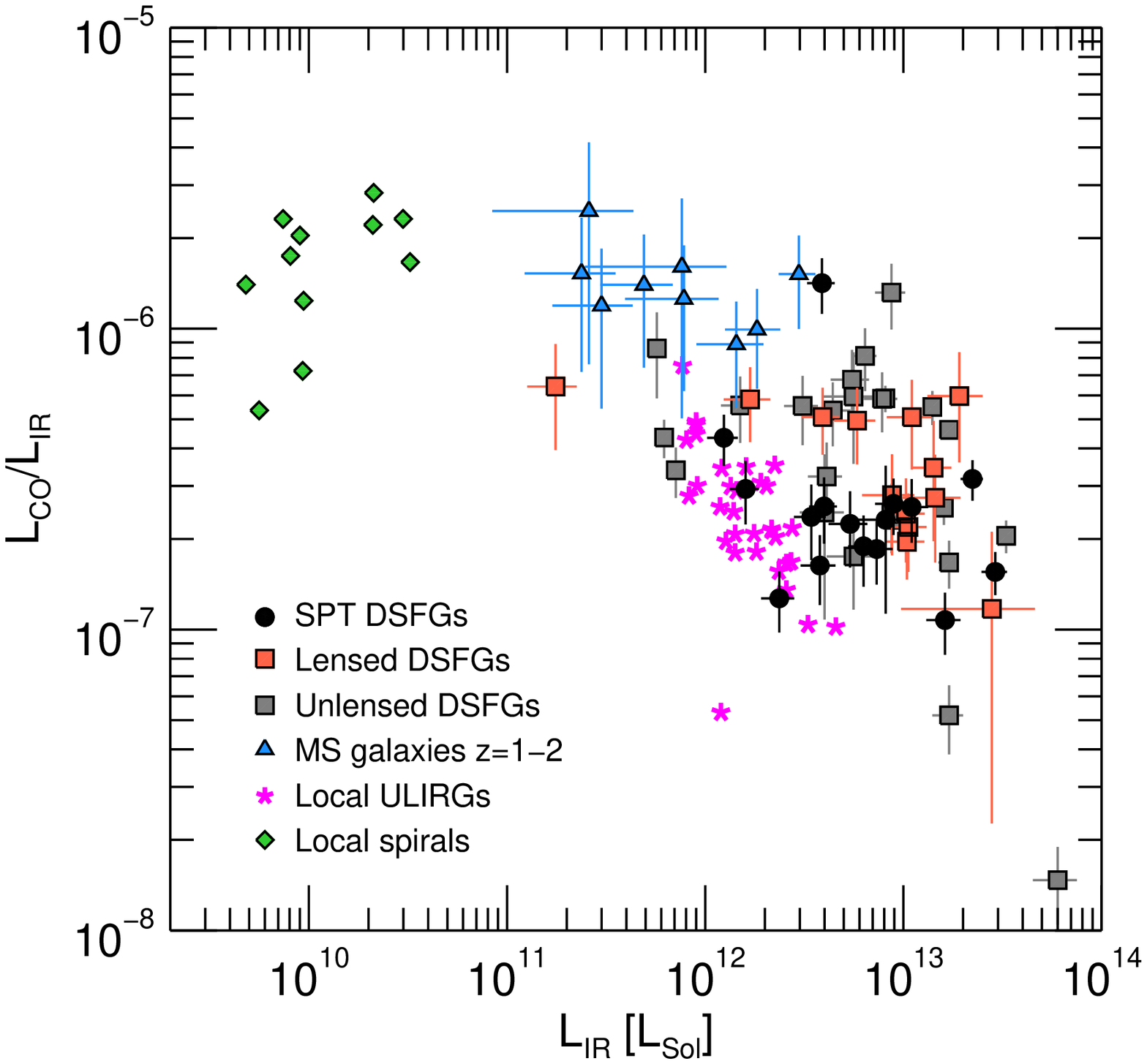}
\caption{$L_{\rm IR}$ versus $L'_{\rm CO}$ ({\it left}) and $L_{\rm CO}/L_{\rm IR}$ vs $L_{\rm IR}$ ({\it right}) for different galaxy populations that have been detected in low-J CO line emission ($J<3$). Gravitationally lensed sources are corrected for magnification. The SPT DSFGs are shown as black filled circles. Orange squares show measurements of gravitationally lensed DSFGs that were discovered in various {\it Herschel} surveys \citep{ivison10, lestrade11, swinbank10,harris10, decarli12, harris12, fu12}, while grey squares show unlensed DSFGs \citep{riechers11, ivison11, ivison13, frayer08, thomson12, carilli11, hodge13, bothwell13, walter12, combes12, coppin10, debreuck14, riechers11b}, respectively. Blue triangles show measurements of massive disk galaxies at $z>0.4$ \citep{geach11,daddi10, magnelli12}. Green squares show measurements of local spiral galaxies \citep{leroy08}, and magenta stars represent local ULIRGs \citep{solomon97}. Also shown on the left plot are representative fits to local spiral and disk galaxies at high redshift (grey dashed line), log$_{10} L_{\rm IR} = 1.12\ {\rm log}_{10} L'_{\rm CO} +0.5$, for guidance. The dotted line shows the same line, with a factor $+0.5$ added. }
\label{fig:lco_lir}
\end{figure*}

The CO line profiles obtained for our sources allow us to estimate the dynamical mass contained within the CO emitting region. Unfortunately, our CO observations do not resolve the sources spatially, and therefore we do not have information about their resolved CO kinematics. However, the reconstruction of the ALMA 870$\mu$m maps allow us to constrain the size and geometry of the dust emission  (we do not use the CO sizes derived above). Our lens reconstruction indicates compact sources, with typical effective radii $<2$kpc (Fig. \ref{fig:sizes}) and evidence of double component sources in a few cases (Spilker et al., in prep). 

For simplicity, we estimate the dynamical masses of the SPT DSFGs from equation 10 in \citet{bothwell13}, assuming that the gas is distributed in a virialized spherical system,  e.g. a compact starburst. This avoids prior knowledge of the source orientation needed for the computation of the dynamical mass in a disk geometry. We also assume that their molecular gas and dust have the same distribution and extension, and thus the source radius is derived from lens modelling. Using the dust sizes derived from lens models constitute the best assumption in our case with the current data in hand. 

The virialized spherical geometry dynamical mass estimator will yield a mass $4.4\times$ larger than the one obtained by assuming a disk-like gas distribution for the same source size, CO linewidth and an average inclination parameter. However, for a disk-like molecular gas distribution, it would be more reasonable to assume a larger $R$ ($\sim5$ kpc as found by Ivison et al.), yielding a dynamical mass estimate which is similar to the one obtained for a compact gas distribution. The obtained dynamical masses, for a virialised compact gas distribution, are listed in Table \ref{table:properties3}.

\subsection{Gas masses}

One of the most important applications of low-J CO line flux measurements is that they provide an estimate of the mass of molecular gas in a galaxy. The molecular gas mass can be computed from the CO(1--0) line luminosity using the relation $M_{\rm gas}=\alpha_{\rm CO} L'_{\rm CO}$, where $\alpha_{\rm CO}$ is the gas mass to CO luminosity conversion factor. The actual value of $\alpha_{\rm CO}$ has been found to depend on several parameters of the host galaxies, such as metallicity or environment \citep[see ][ and references therein]{bolatto13}. A significant trend was found where galaxies with lower metallicities show higher $\alpha_{\rm CO}$ values \citep[e.g.,][]{wilson95, boselli02,leroy11, schruba12, genzel12}. A similar dependency is found with morphology, where compact starbursts show $\alpha_{\rm CO}\sim0.8$ K km s$^{-1}$ pc$^2$ \citep{downes98}, whereas more extended disks, such as the Milky Way, show $4-5\times$ higher $\alpha_{\rm CO}$.

An increasing number of galaxies at high-redshift ($z>1$) have an independent $\alpha_{\rm CO}$ estimate and current values range between $0.5-5$ \citep{weiss07,tacconi08, daddi10,ivison11, magdis11, swinbank11, fu12, hodge12, magdis12, magnelli12, walter12, deane13b, fu13, hodge13, ivison13, messias14, spilker15}. Typically, studies of DSFGs have used $\alpha_{\rm CO}=0.8$ K km s$^{-1}$ pc$^2$ as found for local ULIRGs \citep{downes98} under the implicit assumption that they resemble compact starbursts at high redshift \citep{bothwell13}. However, recent studies have shown that this assumption might not always be correct in individual cases \citep{hodge12}. For consistency with other DSFG studies, we adopt $\alpha_{\rm CO}=0.8$ to obtain the gas masses listed in Table \ref{table:properties3}. In Section \ref{sect_gas}, we show that the properties of our sources agree with this assumption. Unless otherwise stated, we use $\alpha_{\rm CO}$ in units K km s$^{-1}$ pc$^2$.

\section{Discussion}

\label{sect:discussion}

\subsection{Star formation efficiencies and gas depletion}

\begin{figure}
\centering
\includegraphics[scale=0.45]{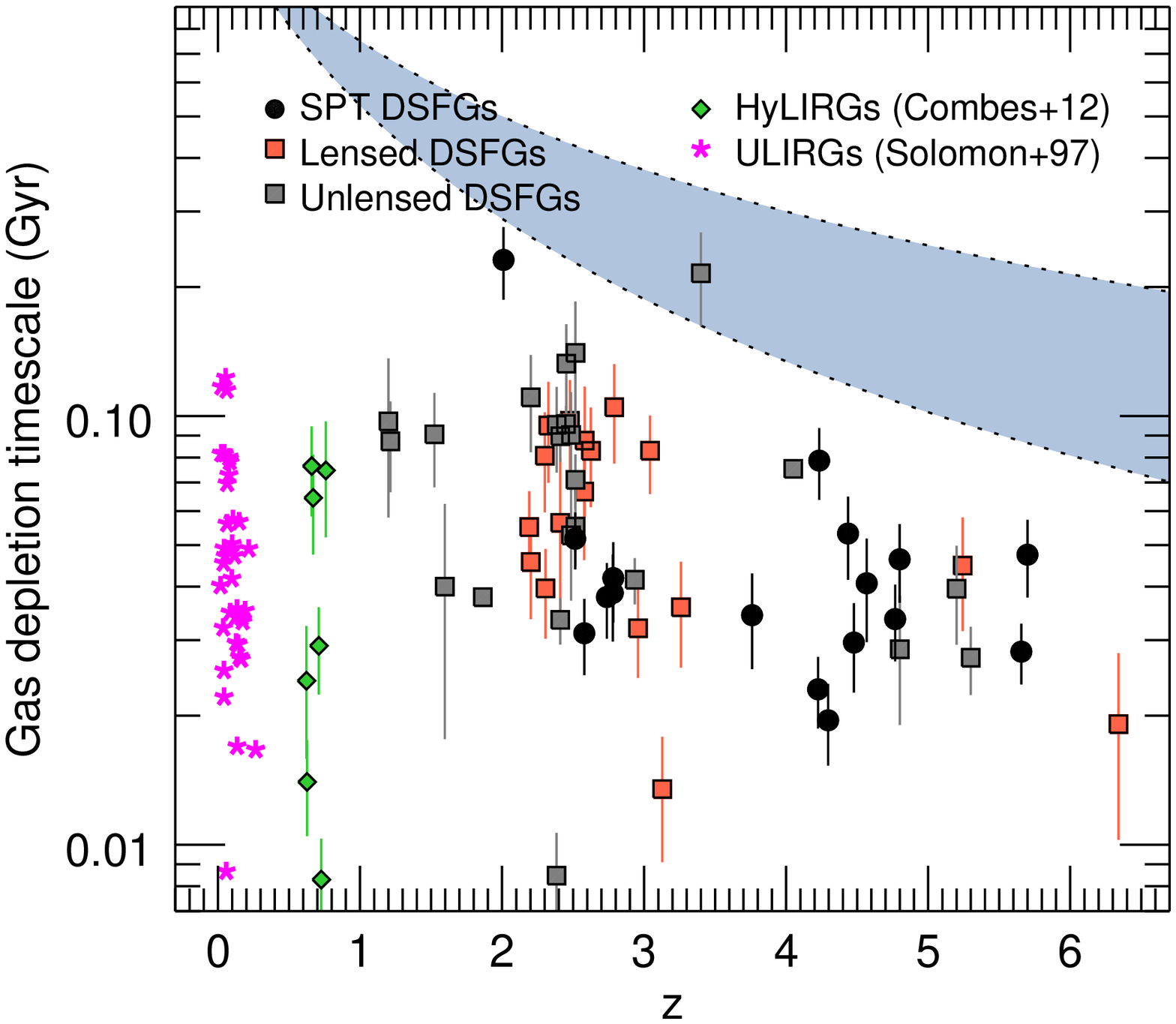}
\caption{\label{fig:tdep} Gas depletion timescale ($t_{\rm dep}$) versus redshift ($z$) for the SPT DSFGs (black circles), compared with other DSFGs from the literature: local ULIRGs \citep[magenta stars; ][]{solomon97}, HyLIRGs \citep[green diamonds; ][]{combes12}, unlensed DSFGs \citep[grey squares; ][]{ivison10, ivison11, ivison13, coppin10, debreuck14, frayer08,  carilli11, hodge13, riechers11, thomson12, walter12, bothwell13, riechers14} and lensed DSFGs \citep[orange circles; ][]{ivison10, lestrade11, swinbank10,harris10, decarli12, harris12, fu12}. The shaded regions enclose the curves $1.5\times(1+z)^{1.0-1.5}$, which is where MS galaxies at $z<3$ are observed to reside \citep{saintonge13}. Measurements for lensed sources assume no differential lensing.}
\end{figure}

The star formation efficiency (SFE) can be defined as the ratio between the SFR and and the molecular gas mass, SFE $={\rm SFR}/M_{\rm gas}$ in units yr$^{-1}$. The inverse of this quantity has been termed the gas depletion timescale, which corresponds to the amount of time required to exhaust all the reservoir of molecular gas at the current rate of star formation, with ($t_{\rm dep}=M_{\rm gas}/SFR$). Typically, given the lack of knowledge about the $\alpha_{\rm CO}$ factor, and thereby intrinsic uncertainty in the estimation of the gas mass, a proxy indicator to the actual SFE has been defined as $L_{\rm IR}/L'_{\rm CO}$ in units $L_{\odot}$ (K km s$^{-1}$ pc$^2)^{-1}$. 

Figure \ref{fig:lco_lir} shows the observed IR and CO luminosities for the SPT DSFGs compared to other galaxy samples that have comparable observations of low-J CO lines. The SPT DSFGs show a SFE range of $\sim100-500$ $L_{\odot}$ (K km s$^{-1}$ pc$^2)^{-1}$, which is higher than the range occupied by local spiral galaxies and $z\sim0.4-2$ main sequence galaxies of $\sim20-100$ (same units). It is interesting to note from the comparison between $L_{\rm IR}$ and $L'_{\rm CO}$ shown in Fig. \ref{fig:lco_lir}{\it -left} that with the exception of one source all SPT DSFGs align with the dotted line, which is representative of higher SFE and consistent with local ULIRGs. Figure \ref{fig:lco_lir}{\it -right} also illustrates this point, and suggests a ``deficit'' of CO luminosity output with respect to $L_{\rm IR}$ for more luminous systems (lensed DSFGs are still more luminous than local ULIRGs after magnification correction). While the reason for this is likely that more luminous systems are producing stars at larger rates and therefore consuming the gas fuel faster, it is also possible that in lensed DSFGs the IR emission is being differentially magnified with respect to the cold molecular gas traced by the low-J CO emission, since the IR emission is tracing the in-situ star forming regions whereas the CO could be tracing more extended regions within each galaxy. This has been suggested by recent high resolution CO observations of two SPT DSFGs \citep{spilker15}. 

Recent observations of main-sequence galaxies have presented evidence for a redshift dependence of the gas depletion timescale with redshift out to $z=2$, with the form $\propto(1+z)^{-1}$ \citep{tacconi13, saintonge13, genzel15}. These observations are supported by theoretical models that seek to explain the evolution of typical star forming galaxies using an equilibrium framework, where galaxies have a steady cosmological gas inflow supply that allows them to maintain significant star formation activity over several Gyrs. Such models show that the gas depletion timescales for MS galaxies should evolve as $\propto(1+z)^{-1.5}$ \citep{dutton10, dave12}. With this form, MS galaxies should have roughly constant gas depletion timescales with redshift, at $z>3$. Note that recent results found by \citet{genzel15}, show a weaker dependency of $t_{\rm dep}$ with redshift ($\propto(1+z)^{-0.3}$) for MS galaxies. While this behaviour has been observed in MS galaxies, both lensed and unlensed \citep{saintonge13}, it is predicted that galaxies undergoing major mergers should have scattered properties, as they are out of an equilibrium state, and hence should not comply to the aforementioned $t_{\rm dep}$ evolution with redshift. 

In Fig. \ref{fig:tdep} we explore the evolution of the gas depletion timescale with redshift for the SPT DSFGs compared to other DSFGs from the literature with low-J CO observations. We assume a conversion between the SFR and IR luminosity of SFR$=10^{-10} L_{\rm IR}$ for a Chabrier initial mass function \citep{chabrier03}, and $M_{\rm gas}=\alpha_{\rm CO} L'_{\rm CO}$ and $\alpha_{\rm CO}=0.8$. Neither parameter, $t_{\rm dep}$ nor $z$, depends on the lensing magnification. We assume no differential magnification between the CO and dust emission. We find that the gas depletion timescales for SPT DSFGs range between $\sim10-100$ Myr. Furthermore, the data show that the gas depletion timescales are fairly homogeneous and independent of redshift, suggesting little evolution of this parameter with cosmic time. In all the sources shown in Fig. \ref{fig:tdep}, we are assuming $\alpha_{\rm CO}=0.8$. If we were to scale the gas masses to $\alpha_{\rm CO}=4.6$, which is not supported by our observations (see below), the data points of the high-redshift objects would lie within the shaded region predicted for MS galaxies, but with a large scatter. 

\subsection{$\alpha_{\rm CO}$ conversion factor}
\label{sect_gas}

\begin{figure}
\centering
\includegraphics[scale=0.45]{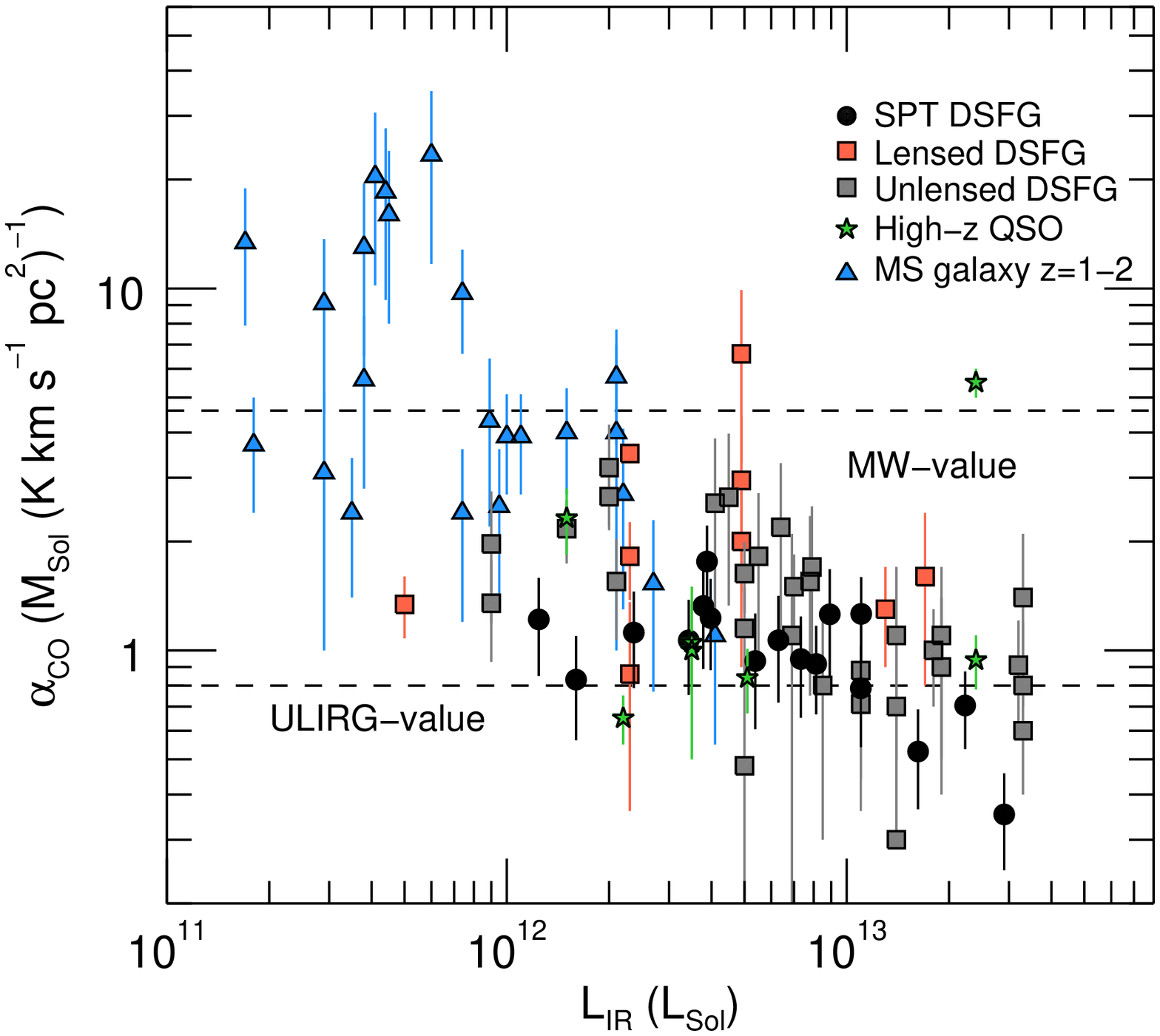}
\caption{$\alpha_{\rm CO}$ factor versus IR luminosity for SPT SMGs computed from the comparison between the dust-derived M$_{\rm ISM, dust}$ and $L'_{\rm CO}$. Also shown are literature compiled values for lensed and unlensed DSFGs, high-redshift quasars and main sequence galaxies \citep{weiss07, daddi10,ivison11, magdis11, swinbank11, fu12, hodge12, magdis12, magnelli12, walter11, walter12, deane13b, fu13, hodge13, alaghbandzadeh13,ivison13, messias14, spilker15}.}\label{fig:alpha}
\end{figure}

A different approach to obtain molecular gas masses for galaxies is to use the dust mass as a proxy for the ISM mass content \citep{leroy11, magdis11, magnelli12, scoville14, genzel15}. The dust masses can be computed from dust model fits to the IR/submm photometry data. Thereby, under a reasonable assumption of the gas-to-dust ratio ($\delta_{\rm GDR}$), and that the ISM is all molecular, gas masses can be extracted using $M_{\rm gas}=\delta_{\rm GDR} M_{\rm dust}$. The advantage of this approach, for high-redshift samples in particular, is the wealth of IR data available for these samples which allows for the computation of accurate dust models (e.g. from Herschel surveys). This method can also be observationally less expensive than measuring CO line fluxes. 

By comparing the molecular gas masses derived from dust models with the measured CO luminosities one can provide an estimate of the $\alpha_{\rm CO}$ conversion factor for individual galaxies from $\alpha_{\rm CO}=M_{\rm gas}/L'_{\rm CO}$. In Fig. \ref{fig:alpha} we show the $\alpha_{\rm CO}$ values for the SPT DSFGs derived using $\delta_{\rm GDR}=100$ \citep{sandstrom13}, and assuming no differential magnification between the CO and dust emission. The IR luminosities of the SPT DSFGs have been corrected by lensing magnification by either using the real $\mu$ value or assuming the average magnification of the sample, $\mu=10$, when individual values were not available. For comparison, measurements of $\alpha_{\rm CO}$ in high-redshift galaxies are also shown. These have been computed using a variety of methods including dynamical mass estimates, CO luminosity surface density and gas-to-dust mass ratios. Our results indicate that most SPT DSFGs have values of $\alpha_{\rm CO}\sim1$, consistent with other similarly luminous DSFGs and high-redshift QSOs from the literature. Conversely typical $z\sim2$ MS galaxies show larger values of $\alpha_{\rm CO}$ suggesting a different nature compared to SPT DSFGs. This may also indicate a selection effect, where we have not observed faint enough SPT sources to populate that region of the diagram yet ($L_{\rm IR}$ cutoff).

This method for computing $\alpha_{\rm CO}$ values has important caveats that need to be mentioned. First, this method relies on the simplistic assumption that the dust can be modelled by a single dust temperature. The dust SED may have multiple dust components contributions throughout the galaxy, which are not accounted for in these models. This yields an underestimation of the dust masses and thus our $\alpha_{\rm CO}$ estimates will be biased low. Second, if the ISM in these galaxies is highly turbulent, a significant fraction of the gas will have high densities. The $\alpha_{\rm CO}$ factor of such dense gas components will be higher than that from the diffuse emission that dominates the low-J CO lines \citep{papadopoulos12b}. Thus, the measured $\alpha_{\rm CO}$ from low-J CO observations will be an average of all components, which will not be representative of the high $\alpha_{\rm CO}$ for dense gas \citep[see e.g.,][]{weiss07}. This is exemplified by the two different determinations of $\alpha_{\rm CO}$ for the quasar host galaxy APM08279+5255. Measurements based on the CO(1--0) line \citep{weiss07}, here dominated by dense gas, lead to $\alpha_{\rm CO}\sim5.0$ while and [CI] emission line measurements \citep{walter11}, related to diffuse regions, yield values closer to $\sim1$ (see Fig. 7).

These two caveats are applicable to any method aiming to compute the $\alpha_{\rm CO}$ based on low-J CO in DSFGs. Finally, and most importantly, the $\delta_{\rm GDR}$ value can vary strongly as a function of metallicity scaling to values up to $\sim1000$ \citep{sandstrom13,remyruyer14}. Given the lack of metallicity estimates for our sample, we adopted a fixed value of $\delta_{\rm GDR}=100$ for all SPT DSFGs close to the average of $\delta_{\rm GDR}=72$ ($\sim0.2$ dex scatter) determined for a large sample of local star forming galaxies with solar metallicities \citep{sandstrom13}. Since it would not be surprising to find low metallicities in high redshift galaxies, it is possible that at least a fraction of our sample has high $\alpha_{\rm CO}$ values. Conversely, we should also consider that the increased star formation activity in DSFGs, evidenced by the large intrinsic SFRs, would cause an enrichment of the galaxy's ISM, lowering the gas content and increasing the metallicity, thereby bringing down the $\delta_{\rm GDR}$ and $\alpha_{\rm CO}$ values \citep{dave12}. 
\subsection{The evolution of gas fraction at $z>2$}

If we assume that the star formation is a process that mainly depends on the amount of available gas, then the evolution of the star formation in the Universe is a direct consequence of the evolution of molecular gas supply \citep{aravena12,bothwell13,walter14}. Recent studies have shown that there is a strong evolution in the molecular gas fraction with increasing redshift \citep{tacconi10, tacconi13, saintonge13}. Current measurements indicate that main-sequence, normal star-forming galaxies at $z\sim2$ tend to have larger reservoirs of molecular gas with respect to the bulk amount of baryonic material than they have in the local Universe, with the molecular gas fraction defined as $f_{\rm gas}=M_{\rm gas}/(M_{\rm stars}+M_{\rm gas})$ and $M_{\rm stars}$ as the stellar mass \citep[e.g.,][]{genzel15}. Using a sample of high-redshift DSFGs, \citet{bothwell13} suggested that the molecular gas fraction in galaxies tends to stay constant at $z>2$, which is well in line with semi-analytic predictions of galaxy formation \citep{lagos11}. 

In this section, we compute the molecular gas fraction for the SPT DFSGs and compare it with other galaxy populations from the literature. Since we do not have stellar mass measurements for all our sample, we instead compute the molecular gas fraction as $f_{\rm gas}= M_{\rm gas}/M_{\rm dyn}$, where $M_{\rm dyn}$ is the dynamical mass. This assumes that the ISM is molecular dominated. 

\begin{figure}
\centering
\includegraphics[scale=0.45]{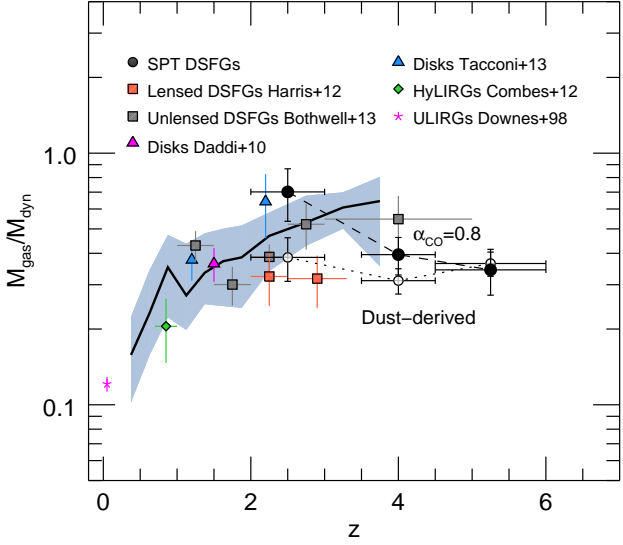}
\caption{Evolution of gas fractions for DSFGs, computed as $M_{\rm gas}/M_{\rm dyn}$. The solid black circles show the gas fraction estimates for SPT DSFGs, where the gas masses were computed from CO luminosities assuming $\alpha_{\rm CO}=0.8$. The empty circles show the gas fractions for SPT DSFGs, with the gas masses derived from dust models.  Also shown are the gas fraction for other galaxy populations: local ULIRGs \citep[magenta star; ][]{downes98}; HyLIRGs \citep[green diamonds; ][]{combes12}; MS galaxies at $z\sim1.5$ \citep[magenta; ][]{daddi10}, and at $z\sim1.2$ and 2.2 \citep[blue; ][]{tacconi13}; unlensed DSFGs \citep[grey squares; ][]{bothwell13}; lensed DSFGs \citep[orange squares][]{harris12}. The gray shaded area shows the average gas fraction for massive star forming galaxies from models of \citet{bethermin15a}. All gravitationally lensed objects have been corrected for the magnification.}\label{fig:fgas}
\end{figure}

Figure \ref{fig:fgas} shows the molecular gas fraction computed in this way for the SPT DSFGs as a function of redshift. For the SPT DSFGs, the gas fractions are provided using two different estimates of the gas mass, one using the CO luminosities and assuming $\alpha_{\rm CO}=0.8$ (shown by solid black circles) and another using the dust masses as explained in the previous section. For comparison, we also show the gas fraction computed in the same manner for other samples in the literature, including ULIRGs \citep[][]{downes98}, HyLIRGs \citep[][]{combes12}, MS galaxies at $z\sim1-2$ \citep{daddi10, tacconi10, tacconi13}, unlensed DSFGs at $z\sim1-4$ \citep{bothwell13} and lensed DSFGs at $z\sim2$ \citep{harris12}. For MS galaxies, $M_{\rm dyn}$ was computed assuming a disk like geometry. Where available, $M_{\rm dyn}$ estimates were obtained directly from resolved CO images \citep{downes98, daddi10}, as they provide more accurate estimates. For HyLIRGs and unlensed DSFGs, dynamical masses were computed using a virialized spherical geometry as explained in Section \ref{sect:mdyn}, assuming a 1 kpc source radius. For lensed DSFGs, we only take into account those with measured magnification factors, and compute $M_{\rm dyn}$ using a spherical geometry and a source radius obtained from the lens model \citep[][Spilker et al., in prep]{bussmann13}.

The shaded area shows the evolution of the average gas fraction ($f_{\rm gas}= M_{\rm gas}/M_{\rm dyn}$) computed by \citet{bethermin15a}. These measurements are based on stacking analysis of the SEDs of a sample of massive star forming galaxies ($M_{\rm stars}>3\times10^{10}\ M_\odot$), from which they deduce average dust masses, and thereby converting to gas masses using a local calibration of the metallicity dependent gas to dust ratio ($\delta_{\rm GDR}$). It should be noted that different $\alpha_{\rm CO}$ values have been assumed by the different studies, with $\alpha_{\rm CO}=0.8$ for ULIRGs, HyLIRGs and DSFGs, and a Milky Way-like $\alpha_{\rm CO}$ for MS galaxies at high redshift. 

Using this metric, Fig. \ref{fig:fgas} shows an overall decrease in the gas fraction from $z\sim1$ to 0, particularly seen in HyLIRGs. At higher redshifts the molecular gas fraction appear to stay almost constant within the uncertainties out to $z\sim5$. This supports previous findings by \citet{bothwell13} and agrees with the overall trend presented by Bethermin et al. out to $z=3.5$. Similarly, this seems to be in line with the relative homogeneity of the gas depletion timescales with redshift. However, there is significant scatter in individual $M_{\rm dyn}$ and $t_{\rm dep}$ values, which may be explained by the starbursting nature of the sample and by uncertainties in the $M_{\rm dyn}$ estimates. 

SPT DSFGs appear to have average gas fractions of $\sim0.5$, similar to what is found in main-sequence galaxies at $z=1-2$.  The comparison between the gas and dynamical masses can provide constraints on $\alpha_{\rm CO}$. Assuming $M_{\rm dyn}>M_{\rm gas}=\alpha_{\rm CO} L'_{\rm CO}$, yields $\alpha_{\rm CO}<M_{\rm dyn}/L'_{\rm CO}$. For the measured CO luminosity and linewidth along with a maximum effective radius from the dust-derived lens models ($2$ kpc), a limit $\alpha_{\rm CO, lim}$. Individual limits computed thereby are listed in Table 3. In this computation we are assuming that all the mass is in the form of molecular gas; however, it is still possible that the gas is extended outside the radius defined by dust. This is not possible to quantify without higher resolution CO imaging. Overall, these limits imply that a major fraction of the SPT DSFGs must have low $\alpha_{\rm CO}$, $<4$, however, it leaves room for up to $\sim30\%$ of the sources to have larger values.

\section{Conclusions}
\label{sect:conclusions}
In this paper, we have presented low-J CO observations of a sample of 17 SPT DSFGs that have precise redshift measurements from ALMA CO-based spectroscopy. Our main results are the following:

\begin{itemize}
\item We detect CO line emission in 17 gravitationally lensed DSFGs from the SPT millimetre survey. The obtained CO luminosities imply molecular gas masses in the range  $(1.3-6.3)\times10^{10}\ (\alpha_{\rm CO}/0.8) (\mu/10)^{-1}\ M_{\odot}$. Comparison with the total IR luminosities indicate short gas depletion timescales ($<100$ Myr) or high star formation efficiencies, comparable to that of local ULIRGs.

\item Using our CO measurements and accurate lens models for our sample, we quantified the ability to find lensing magnification factors based on the measured CO luminosities and line-widths. We find that this method is highly uncertain, typically finding a magnification $\mu$ to within 50\% uncertainty, in only 33\% of the cases. 

\item Based on the dust masses computed from multi-wavelength SED fitting, we compute gas masses assuming a typical gas-to-dust mass ratio of 100. Comparison of this dust-derived gas mass estimate with the CO luminosities result in low $\alpha_{\rm CO}$ factors for most sources in our sample, with typical $\alpha_{\rm CO}\sim1$. Such values are similar to that found in the most luminous objects ($>10^{12}\ L_\odot$) at high-redshift, and are consistent with the values found for local ULIRGs. Several caveats in the computation of this parameter are presented.

\item We use the dynamical and gas masses computed from our CO measurements to constrain the average gas fraction ($M_{\rm gas}/M_{\rm dyn}$) as a function of redshift. We find that our results are consistent with previous studies of DSFGs that suggest that the gas fraction stays almost constant at $z>2$.

\end{itemize}

The CO observations presented in this work support the finding that SPT DSFGs are undergoing an active, short-lived starburst episode. These indicate large reservoirs of molecular gas that is however not enough to sustain the star formation activity for more than a few 100 Myr (the case even if we assume a Milky Way -like $\alpha_{\rm CO}$ factor). This is reflected in Fig. \ref{fig:lco_lir}, where SPT DSFGs appear located 0.5 dex above the sequence occupied by distant main-sequence galaxies and local, normal spirals \citep{daddi10}. Further evidence comes from the derived $\alpha_{\rm CO}$ conversion factor which implies a value of $\sim1$ for most individual sources, despite the several caveats on this computation. Besides suggesting a uniformity in the ISM conditions in our sample, it also suggest that the ISM has similar conditions to those found in local ULIRGs, for which a typical value of $\alpha_{\rm CO}=0.8$ is found \citep{downes98}. The final piece of supporting evidence corresponds to the recent results of \citet{spilker15}. They perform a source-plane reconstruction of the distribution of molecular gas and star formation in two SPT DSFGs, SPT0538-50 and SPT0346-52 at $z=2.8$ and $z=5.7$, respectively, both included in the present sample. In the first case, their results suggest a pair of merging galaxies with extended distribution of molecular gas, while in the second case, it shows disturbed dynamics. 

Even though the current evidence therefore points toward the conclusion that a dominant fraction of SPT DSFGs are starbursts, we cannot discard the possibility that some SPT DSFGs are driven by cold accretion given the increasing number of DSFGs at high redshift with evident disk-like morphology \citep{swinbank11,hodge12, carniani13,debreuck14}.

\section*{Acknowledgements}

M.A. acknowledges partial support from FONDECYT through grant 1140099. J.S.S., D.P.M., and J.D.V. acknowledge support from the U.S. National Science Foundation under grant No. AST-1312950. JSS acknowledges support through award SOSPA1-006 from the NRAO. The Australia Telescope is funded by the Commonwealth of Australia for operation as a National Facility managed by CSIRO. The SPT is supported by the National Science Foundation through grant ANT-0638937, with partial support through PHY-1125897, the Kavli Foundation and the Gordon and Betty Moore Foundation.

%%%%%%%%%%%%%%%%%%%%%%%%%%%%%%%%%%%%%%%%%%%%%%%%%%



\bibliographystyle{mnras}
\bibliography{../../bibtex/spt_smg}

% Don't change these lines
\bsp	% typesetting comment
\label{lastpage}
\end{document}